\documentclass[11pt,a4paper]{article}
\usepackage{jcappub}
\usepackage{graphicx}
\usepackage{enumerate}
\usepackage{tikz}
\usepackage{amsmath}
\usepackage{amssymb}
\usepackage{commath}
\usepackage{bbold}

\usepackage{dsfont}
\usepackage{hyperref}
\usetikzlibrary{arrows,positioning}

\newcommand{\vect}[1]{\boldsymbol{\mathbf{#1}}}

\newcommand{\fex}{\textit{e.g.}}
\newcommand{\ie}{\textit{i.e.}}

\newcommand{\tobs}{T_{\rm obs}}
\newcommand{\epo}{\mathcal{E}}
\newcommand{\aeff}{A_{\rm eff}}

\begin{document}

\title{A Fresh Approach to Forecasting in Astroparticle Physics and Dark Matter Searches}

\author{Thomas D. P. Edwards}
\emailAdd{t.d.p.edwards@uva.nl}
\author{and Christoph Weniger}
\emailAdd{c.weniger@uva.nl}

\affiliation{Gravitation Astroparticle Physics Amsterdam (GRAPPA), Institute for Theoretical Physics Amsterdam and Delta Institute for Theoretical Physics, University of Amsterdam, Science Park 904, 1090 GL Amsterdam, The Netherlands}

\abstract{
  We present a toolbox of new techniques and concepts for the efficient forecasting of experimental sensitivities.  These are applicable to a large range of scenarios in (astro-)particle physics, and based on the Fisher information formalism.  Fisher information provides an answer to the question `what is the \emph{maximum} extractable information from a given observation?'.  It is a common tool for the forecasting of experimental sensitivities in many branches of science, but rarely used in astroparticle physics or searches for particle dark matter.  After briefly reviewing the Fisher information matrix of general Poisson likelihoods, we propose very compact expressions for estimating expected exclusion and discovery limits (`equivalent counts method').  We demonstrate by comparison with Monte Carlo results that they remain surprisingly accurate even deep in the Poisson regime.  We show how correlated background systematics can be efficiently accounted for by a treatment based on Gaussian random fields.  Finally, we introduce the novel concept of Fisher \emph{information flux}.  It can be thought of as a generalization of the commonly used signal-to-noise ratio, while accounting for the non-local properties and saturation effects of background and instrumental uncertainties.  It is a powerful and flexible tool ready to be used as core concept for informed strategy development in astroparticle physics and searches for particle dark matter.
}


\arxivnumber{1704.05458}

\maketitle

\section{Introduction}

Progress in astroparticle physics and dark matter (DM) searches is driven by the comparison of theoretical models with experimental data. During this process, estimating the sensitivity of existing or future experiments for the detection of astrophysical or new physics signals is a ubiquitous task, usually requiring the calculation of the expected exclusion and discovery limits.  It is done both by phenomenologists who are interested in observational prospects of their theoretical models, as well as experimentalists who are interested in optimizing experimental design.  As such, efficient and informative forecasting plays a central role in shaping the development of the field.  It is rather common in the (astro-)particle physics community that forecasting is done in the framework of Frequentist statistics.  Acceptance of the signal+background hypothesis, $H_1$, and rejection of the background-only hypothesis, $H_0$, requires that some test statistic (TS) exceeds a predefined threshold, which depends both on the aspired significance level of the detection as well as the probability distribution function (PDF) of the TS under $H_0$~\cite{Cowan:1998ji, James:2006zz}.  The Frequentist method (in contrast to Bayesian techniques) has the advantage of known false positive rates for hypothesis testing, and known coverage for upper limits. These features can be especially desirable when prior knowledge about the (non-)realization of a theoretical models in Nature is poor.

One of the most commonly used TSs (used for parameter regression, calculation of confidence intervals, and the goodness-of-fit) is the \emph{chi-squared}. Its application is limited to binned data with errors that are approximately normal distributed, which is often realized in the large-sample limit.  Asymptotic formulae for its statistical properties are available~\cite{Cowan:2010js}.  In the small sample limit, and generally for more complicated likelihood functions, the maximum likelihood ratio (MLR)~\cite{Cowan:2010js} is a very common TS.  Again, in the large sample limit, asymptotic distributions are available~\cite{Wilks:1938a, Chernoff:1954a}.  In the small-sample limit, however, its statistical properties have to be inferred from Monte Carlo (MC) simulations.  
  
\medskip

When estimating experimental sensitivities, one general question that arises is `\emph{What is the maximum information that can be in principle extracted from a given observation?}'.  Information gain here corresponds to the reduction of the uncertainty associated to the model parameters of interest.  The famous Cram\'er-Rao bound~\cite[e.g.][]{James:2006zz} provides a general way to derive, for a given experimental setup, a lower bound on the achievable variance of any unbiased model parameter estimator, which holds for \emph{any} statistical method employed to analyse the data.  This bound corresponds to an upper limit on the achievable information gain.  The Cram\'er-Rao bound is based on the \emph{Fisher information matrix}, which quantifies how `sharply peaked' the likelihood function describing the observational data is around its maximum value.

Forecasting instrumental sensitivities with Fisher information is rather common, \fex, in cosmology.  It rests on the assumption that estimators which saturate the Cram\'er-Rao bound are available, and that these estimators approximately follow a multi-variate normal distribution.  In the large-sample limit, this is indeed often (but not always) the case.  The important limitations of this approach were pointed out many times~\cite[\fex,][]{Wolz:2012sr, Vallisneri:2007ev}, as well as proposals to extend the framework to account for, \fex, non-Gaussian effects~\cite{Sellentin:2014zta} (for a recent review see Ref.~\cite{Heavens:2016slh}).

The Fisher information matrix has an impressive range of attractive properties that -- if used wisely -- can significantly ease the life of anybody interested in performing forecasting, from simple tasks to problems with many experiments, targets and a high-dimensional parameter space.  It can be quickly calculated, it is additive, allows for efficient handling of nuisance parameters, and it is at the root of the powerful concept of information geometry~\cite[\fex,][]{Brehmer:2016nyr} (and, as we will show in this work, \emph{information flux}). Fisher forecasting is largely unused in the astroparticle physics and DM communities however there are some exceptions, \fex ~\cite{Wolz:2012sr, Camera:2014rja, Bringmann:2016axu, Bartels2017, Charles:2016pgz, Brehmer:2016nyr, Abdo2013}.

\medskip

In this paper, we present an overview of how Fisher information can be used in the context of astroparticle physics and DM searches.  Throughout this paper, we focus our attention on Poisson likelihoods and \emph{additive component models}, where the shape (in for instance energy or positional space) of the model components are fixed and the normalization coefficients are the only regression parameters.  In addition, we will study the impact of additional \emph{external constraints} on the model parameters.  Such constraints can be used to account for various model or instrumental uncertainties.  Focusing on these scenarios allows us an in-depth discussion of the specific capabilities and limitations of Fisher forecasting, while still covering many interesting use cases.

We introduce various (to the best of our knowledge) new prescriptions for the efficient estimate of the expected exclusion and discovery limits that are valid in the small- and large-sample regimes based on the Fisher information matrix.  We compare the accuracy of these prescriptions with results from the MLR method, and for a few simple cases with results from the full Neyman belt construction.  We demonstrate how to incorporate the effect of correlated background systematics.  Lastly, we introduce the new concept of Fisher \emph{information flux}.  It generalizes the commonly used signal-to-noise ratio (SNR), while fully accounting for background and other uncertainties.  We illustrate the power of this new concept in a few simple examples.

Some of the techniques discussed in this paper are extensively used in other fields of science.  Part of the work is inspired by the discussion of information geometry in Ref.~\cite{Brehmer:2016nyr}, and by the notion of effective backgrounds in Refs.~\cite{Albert:2014hwa, Charles:2016pgz}.  We mention explicitly when and why we deviate from notations introduced in these works if we deem this to be necessary.  Some of the methods discussed here have already been applied by some of the present authors in Refs.~\cite{Bringmann:2016axu, Bartels2017}. 

\medskip

The paper is organized as follows:  In Sec.~\ref{sec:fisher}, we introduce and define the Fisher information matrix, additive component models, and equivalent signal and background counts.  In Sec.~\ref{sec:estimates}, we present prescriptions to derive approximate expected exclusion limits and discovery sensitivities from the Fisher information matrix, and study the validity of the results with MC techniques.  In Sec.~\ref{sec:systematics} we show in a few examples how systematic background uncertainties can be incorporated in the sensitivity estimates.  In Sec.~\ref{sec:optimization} we introduce the notion of Fisher information flux, and various connected concepts.  In Sec.~\ref{sec:conclusions} we finally conclude.

In Appendix~\ref{apx:PL} we discuss relevant properties of the Poisson likelihood function and the associated Fisher information.  In Appendix~\ref{apx:UL}, we discuss conventional methods for the calculation of the discovery reach and expected upper limits, that we use for comparison with our techniques.  Finally, in Appendix~\ref{apx:technical} we collect some more technical derivations for results in the paper.

\section{Fisher Information of the Poisson likelihood function}
\label{sec:fisher}

\subsection{The Fisher information matrix}

For any sufficiently regular likelihood function $\mathcal{L}(\mathcal{D}|\vect\theta)$, with $n$ model parameters $\vect\theta \in \mathbb{R}^n$ and data $\mathcal{D}$, the Fisher information matrix is a $n\times n$ matrix that can be defined as
\begin{equation}
  \mathcal{I}_{ij}(\vect\theta) \equiv  \left\langle\left(\frac{\partial
    \ln \mathcal{L}(\mathcal{D}|\vect{\theta})
    }{\partial \theta_i}\right)\left(\frac{\partial
    \ln \mathcal{L}(\mathcal{D}|\vect{\theta})
}{\partial \theta_j}\right) \right\rangle_{\mathcal{D}(\vect\theta)}
 =   -\left\langle\frac{\partial^2
    \ln \mathcal{L}(\mathcal{D}|\vect{\theta})
    }{\partial \theta_i \partial
  \theta_j} \right\rangle_{\mathcal{D}(\vect\theta)}\;,
  \label{eqn:I}
\end{equation}
where the average is taken over multiple realizations of a model with parameters $\vect\theta$.  The second equality holds given some weak regularity conditions \cite{Casella_book}.  The Fisher information matrix quantifies the \emph{maximum precision} at which the model parameters can be inferred from the data.\footnote{Here, we refer to the maximum precision in terms of a Frequentist approach. In the Bayesian case it is possible to gain further insight with the addition of informative priors and/or a hierarchical model structure.} This follows from the Cram\'er-Rao bound (CRB)~\cite{RadhakrishnaRao1945, cramer2016mathematical}, which states that for \textit{any} set of unbiased estimators of the model parameters, $\vect{\hat\theta}$, the inverse of the Fisher information matrix provides a lower limit on its variance. The CRB generalises to the multivariate case to give \cite{Cowan:1998ji},
\begin{equation}
  \text{cov}\left[\hat\theta_i ,\hat\theta_j\right] \equiv \langle (\hat\theta_i-\theta_i) (\hat\theta_j-\theta_j) \rangle_{\mathcal{D}(\vect\theta)} \geq (\mathcal{I}(\vect{\hat\theta})^{-1})_{ij} \;.
  \label{eqn:CRB}
\end{equation}
Estimators that saturate the bound exactly are called `minimum variance'. An estimator is called `asymptotically efficient' when the bound is saturated in the large sample limit.  Note that this bound only holds for \emph{unbiased} estimators, and becomes stronger in the presence of a constant bias~\cite{Cowan:1998ji}.

A widely used estimator is the maximum likelihood estimator (MLE).  The MLE is \emph{exact}, meaning that although it is in general biased it becomes unbiased in the large sample limit.  Furthermore, the MLE is asymptotically efficient.  Although the Fisher information matrix as an estimate for the variance of MLEs becomes exact only in the large-sample limit, we will see below that it remains a powerful tool in the small-sample regime.

\medskip

Searches for new physics in both astrophysics and DM detection experiments are often, at their core, counting experiments. The number of events recorded in a detector is described by the Poisson distribution. We assume here for simplicity that events can be described by one variable, for example photon energy $E$ (the generalization to multiple variables is straightforward and will be used below).  Then, the Poisson log-likelihood can be written as (details can be found in Appendix~\ref{apx:GPL})
\begin{equation}
  \ln \mathcal{L}_\text{pois}(\mathcal{D}|\vect\theta) = \int dE\;
  \left[\mathcal{C}(E)\ln\Phi(E|\vect\theta) - \Phi(E|\vect\theta)\right]\;,
  \label{eqn:lnLp}
\end{equation}
where the integration is done over the energy range of interest, and we dropped terms that do not depend on $\vect\theta$ since they do not affect the rest of the discussion.  Here, $\Phi(E|\vect\theta)$ is the model counts spectrum (with units $1/E$).  Furthermore, we defined the `unbinned' count spectrum
\begin{equation}
  \mathcal{C}(E) = \sum_k \delta_D(E-E_k)\;,
  \label{eqn:unbinned_countmap}
\end{equation}
where $E_k$ is the energy of event $k$, and $\delta_D(E)$ is the Dirac delta function. $\mathcal{C}(E)$ can be interpreted as a counts histogram with zero bin size. It represents a specific realization of a Poisson process with mean $\Phi(E|\vect\theta)$, and has the useful property that $\langle \mathcal{C}(E) \rangle_{\mathcal{D}(\vect\theta)} = \Phi(E|\vect\theta)$.

We note that the above definition resembles a \emph{Poisson point process}, defined here on an interval on the real line that is given by the energy range of interest~\cite{chiu2013stochastic, Coeurjolly:2014a}.  In this context, $\Phi(E|\vect\theta)$ would be referred to as `intensity measure', but we will continue to use here the expressions `model counts spectrum' or, in some 2-dim examples below, `model counts map'.

\medskip

It is straightforward to show that the Fisher information matrix of the above Poisson likelihood is given by
\begin{equation}
  \mathcal{I}_{ij}^\text{pois}(\vect\theta) =
  \int dE\, \frac{\partial\Phi(E|\vect\theta)}{\partial \theta_i}
  \frac1{\Phi(E|\vect\theta)}
  \frac{\partial \Phi(E|\vect\theta)}{\partial \theta_j}
  \;.
  \label{eqn:Fisher}
\end{equation}
This form is fully general and holds for any parametric dependence of $\Phi(E|\vect\theta)$.  The inverse of the Fisher information matrix (irrespectively of whether it is derived from Poisson or other likelihood functions) approximates the expected covariance matrix as shown in Eq.~\eqref{eqn:CRB}.  As mentioned above, this relation holds exactly in the large-sample limit only.

For the diagonal elements of the inverse of the Fisher matrix, we will sometimes use of the notation 
\begin{equation}
  \sigma_i^2(\vect\theta) \equiv (\mathcal{I}(\vect\theta)^{-1})_{ii}\;.
\end{equation}
Furthermore, we note that in general the full Fisher matrix can have, beside the Poisson part that we discussed above, parts that introduce additional constraints on the model parameters (see Sec.~\ref{sec:systematics} below).  In that case, we can write
\begin{equation}
  \mathcal{I}_{ij}(\vect\theta) = \mathcal{I}^\text{pois}_{ij}(\vect\theta) +\mathcal{I}^\text{syst}_{ij}\;,
  \label{eqn:Isplit}
\end{equation}
where we assumed that the systematics term is (approximately) independent of the model parameters.

\subsection{The profiled Fisher information matrix}

Typically, one is only interested in a few `parameters of interest' (PoI), say $ \theta_1, \dots, \theta_k$, while the remaining $n-k$ model parameters are nuisance parameters that parametrize background, instrumental or signal uncertainties.  The canonical method to deal with MLEs of nuisance parameters in a Frequentist approach is to maximise the likelihood function $\mathcal{L}(\mathcal{D}|\vect\theta)$ with respect to the parameters $\theta_{k+1}, \dots, \theta_n$, which gives rise to a `profile likelihood' function that only depends on the $k$ PoI.  This method leaves one with a description of the remaining parameter space whilst accounting for the effects of the $n-k$ nuisance parameters.

To perform the analogous procedure for the Fisher information matrix, we write $ \mathcal{I}_{ij}(\vect\theta)$ in block form,
\begin{equation}
  \mathcal{I}=
  \begin{pmatrix}
    \mathcal{I}_A & \mathcal{I}_C^T \\
    \mathcal{I}_C & \mathcal{I}_B
  \end{pmatrix}\;,
  \label{eqn:block}
\end{equation}
where $\mathcal{I}_A$ is a $k\times k$ matrix that corresponds to the PoI, $\mathcal{I}_B$ is a $(n-k)\times(n-k)$ matrix that corresponds to the nuisance parameters, and $\mathcal{I}_C$ is the mixing between both. We then define the  \textit{profiled Fisher information matrix}, where the nuisance parameters are removed in such a way that the (co-)variance of the PoIs remains unchanged. It is given by\footnote{This result can be readily understood by acknowledging that the upper left $k\times k$ part of the inverse of Eq.~\eqref{eqn:block} is given by $(\mathcal{I}_A-\mathcal{I}_C^T \mathcal{I}_B^{-1} \mathcal{I}_C)^{-1}$.}
\begin{equation}
  \mathcal{\widetilde I}_{A} = \mathcal{I}_A - \mathcal{I}_C^T\, \mathcal{I}_B^{-1} \,\mathcal{I}_C\;.
  \label{eqn:Imarg}
\end{equation}
This expression is general and does not depend on the details of the problem.

\subsection{Additive component models}

To simplify the discussion we will, as mentioned above, assume that the model counts spectrum $\Phi(E|\vect\theta)$ consists of a number of additive components with free normalization $\theta_i$ but fixed shape $\Psi_i(E)$,
\begin{equation}
  \Phi(E|\vect \theta) = \sum_{i=1}^n \theta_i \Psi_i(E)\;,
  \label{eqn:add_model}
\end{equation}
where the sum is taken over the $n$ model components.  The model counts spectra can be in many cases calculated as
\begin{equation}
  \Psi_i(E) = \epo(E)\,I_i(E)\;,
  \label{eqn:def_epo}
\end{equation}
where $\epo(E)$ denotes the exposure (usually effective area or volume times effective observation time) as function of energy, and $I_i(E)$ the differential flux of component $i$.  We furthermore will occasionally use the total flux $I(\vect\theta|\Omega) \equiv \sum_i \theta_i I_i(\Omega)$.  The number of expected counts per component is given by
\begin{equation}
  \lambda_i (\theta_i) \equiv \theta_i \int dE\, \Psi_i(E)\;.
  \label{eqn:lambda}
\end{equation}

For the additive component model, the Fisher information matrix acquires the simple form
\begin{equation}
  \mathcal{I}^\text{pois}_{ij}(\vect\theta) = \int dE \frac{\Psi_i(E) \Psi_j(E)}{\Phi(E|\vect\theta)}\;.
  \label{eqn:Ipsi}
\end{equation}
Note that the model parameters enter here only through the model predictions in the denominator.

\subsection{Equivalent number of signal and background events}
\label{sec:effsb}

In Sec.~\ref{sec:estimates} we will present methods for calculating approximate upper limits and discovery reaches which rely on the definitions of the number of signal and background counts. We refer to the method as the Equivalent Counts Method (ECM) since it is a generalisation of the single binned case where it is clear that the number of signal and background events provides information about (a) the expected significance of the signal, (b) the signal-to-background ratio (SBR) and (c) the sample size and hence the relevance of the discreteness or skewness of the Poisson distribution. The ECM definitions capture the same information but for more general cases i.e. situations with large numbers of bins or even the unbinned case as is presented here. In general, not all signal events $\lambda_i(\theta_i)$ are statistically relevant, since some might overlap with regions containing strong backgrounds, whilst other signal events might reside in regions that are almost background free.  A reasonable definition of equivalent signal and background events should account for this effect.

We propose definitions here for the number of equivalent signal and background counts that are defined \emph{solely} in terms of the Fisher information matrix.  They provide information about the expected signal significance, the SBR and the effective sample size in rather general situations.  These definitions are used below for the calculation of expected exclusion and discovery limits.

\medskip

We define the equivalent signal, $s_i(\vect\theta)$, and background, $b_i(\vect\theta)$, events for any component $i$ implicitly in terms of the SNR,
\begin{equation}
  \text{SNR}_i(\vect\theta) = \frac{s_i^2(\vect\theta)}{s_i(\vect\theta)+b_i(\vect\theta)}\;,
  \label{eqn:snr_connection}
\end{equation}
and in terms of the SBR,
\begin{equation}
  \text{SBR}_i(\vect\theta) = \frac{s_i(\vect\theta)}{b_i(\vect\theta)}\;.
  \label{eqn:sbr_connection}
\end{equation}

The SNR for model component $i$ is given by the corresponding diagonal term of the inverse of the Fisher information matrix, times the factor $\theta_i^2$, which we can write as
\begin{equation}
  \text{SNR}_i(\vect\theta) \equiv \frac{\theta_i^2}{\sigma_i^2(\vect\theta)}\;.
  \label{eqn:snr}
\end{equation}
If systematic (non-Poisson) contributions to the $\mathcal{I}_{ii}(\vect\theta)$ as well as mixing with other components are negligible, $\text{SNR}_i(\vect\theta)$ is essentially given by the corresponding diagonal component of Eq.~\eqref{eqn:Ipsi}.  As expected, this is simply the standard SNR of $\theta_i\Psi_i(E)$ over a fixed background $\Phi(E|\vect\theta)$.  
The definition of the SBR in terms of the Fisher information matrix is less obvious.  We find that the following expression serves the purpose,
\begin{equation}
  \text{SBR}_i(\vect\theta) \equiv \frac{\sigma_i^2(\vect\theta)}{\sigma_i^2(\vect\theta_0)}-1\;.
  \label{eqn:sf}
\end{equation}
Here, we used the definition $\vect\theta_0 \equiv (\theta_1, \dots, \theta_{i-1}, 0, \theta_{i+1}, \dots, \theta_n)^T$, \ie\ it is $\vect\theta$ with the normalization of the signal component set to zero.  This choice will be further justified below.

From the definitions in Eqs.~\eqref{eqn:snr} and~\eqref{eqn:sf}, the equivalent number of signal and background events can be directly obtained using Eqs.~\eqref{eqn:snr_connection} and~\eqref{eqn:sbr_connection}.  They are given by
\begin{equation}
  \label{eqn:si}
  s_i(\vect\theta) = \frac{\theta_i^2}{\sigma_i^2(\vect\theta)- \sigma_i^2(\vect\theta_0)}
  \;,
\end{equation}
and
\begin{equation}
  \label{eqn:bi}
  b_i(\vect\theta) = \frac{\theta_i^2\sigma_i^2(\vect\theta_0)}{(\sigma_i^2(\vect\theta)- \sigma_i^2(\vect\theta_0))^2}
  \;.
\end{equation}

\paragraph*{Discussion.}

Although these definitions are relatively abstract, they have a number of useful properties.  These become particularly clear when mixing between the components is negligible, \ie~$\sigma_i^2(\vect\theta) \simeq 1/\mathcal{I}_{ii}^\text{pois}(\vect\theta)$. If component shapes are very similar, such a situation can be enforced by introducing constraint terms for all non-signal components, in the way that we will discuss below in Sec.~\ref{sec:syst_basics}.  If \emph{component mixing is negligible}, we find:

\textit{(i)} If all $\Psi_i(E)$ are constant, which corresponds to a single-bin scenario, the number of equivalent signal events just equals the total number of signal events, $s_i(\vect\theta) = \lambda_i(\vect\theta)$, and the equivalent background events equals the sum of the events from all other components, $b_i(\vect\theta) = \sum_{j\neq i} \lambda_j(\vect\theta)$.  

\textit{(ii)} In the large-signal limit, here defined as $\theta_i \Psi_i(E) \gg \sum_{j\neq i}\theta_j\Psi_j(E)$, the number of equivalent signal events of component $i$ equals the total number of events in the signal component, $s_i(\vect\theta) \simeq \lambda_i(\theta_i)$.  In other cases, the equivalent number of signal events is in general smaller, $s_i(\vect\theta)\leq \lambda_i(\theta_i)$.

\textit{(iii)} The definition of the SBR for model component $i$ in Eq.~\eqref{eqn:sf} becomes clearer when writing it in terms of the additive components functions $\Psi_i(E)$ and their sum $\Phi(E)$.  In the small-signal regime, $\theta_i\Psi_i(E) \ll \sum_{j\neq i} \theta_j \Psi_j(E)$, this yields
\begin{equation}
  \text{SBR}_i(\vect\theta) \simeq\\
  \left(\int dE\, \frac{\Psi_i(E)^2}{\Phi(E|\vect\theta_0)}\right)^{-1}
  \int dE\, \frac{\Psi_i(E)^2}{\Phi(E|\vect\theta_0)}\; \frac{\theta_i \Psi_i(E)}{\Phi(E|\vect\theta_0)}\;.
  \label{eqn:sf_simple}
\end{equation}
Here, the ratio $\theta_i\Psi_i(E)/\Phi(E)$ can be interpreted as the SBR of component $i$ as function of $E$, and $\Psi_i^2(E)/\Phi(E)$ is proportional to the SNR as function of $E$.  Thus, Eq.~\eqref{eqn:sf_simple} effectively averages the SBR over regions of $E$ where the signal component $\Psi_i(E)$ is most significant w.r.t.~the background $\Phi(E)$.  In other words, regions in $E$ where the background $\Phi(E)$ is intense enough that the signal $\Psi_i(E)$ is swamped do not contribute to Eq.~\eqref{eqn:sf_simple}, even if the number of signal photons in that region is high.  

\textit{(iv)} On the other hand, in the large-signal limit (as defined above), the SBR of component $i$ becomes
\begin{equation}
  \text{SBR}_i(\vect\theta) \simeq
  \left(\int dE\, \Psi_i(E)\right)^{-1}
  \int dE\, \Psi_i(E)\; \frac{\theta_i \Psi_i(E)}{\Phi(E|\vect\theta_0)}\;.
  \label{eqn:sf_approx}
\end{equation}
The weighting by the significance of the signal is now replaced by a weighting over the signal strength.\footnote{Eq.~\eqref{eqn:sf_approx} is underlying the definition of the `effective background', $b_\text{eff}=n_s^2/\text{TS}$, used in Refs.~\cite{Albert:2014hwa, Charles:2016pgz}, if we identify $n_s = \lambda_i$ and $\text{TS}=\text{SNR}_i$ (see Eq.~\eqref{eqn:snr}).  We find that this definition can significantly underestimate the SBR in cases where most of the signal events are located in regions where they are statistically not significant.  This can happen for instance in low-energy tails of a steeply falling astrophysical spectra.  Our definitions based on Eq.~\eqref{eqn:sf} do not exhibit this problem.}

\medskip

In the \emph{general case with mixing} between the components, the above simple analytic expressions do not hold anymore.  If we concentrate on Fisher information matrices of the from Eq.~\eqref{eqn:Isplit}, where the non-Poisson part is independent of the model parameters, one can however still show that $\text{SBR}_i(\vect\theta)\geq0$, which implies the important property that $b_i(\vect\theta)\geq0$ and $s_i(\vect\theta)\geq0$ in all cases.  We confirmed this numerically by calculating $\text{SBR}_i$ for a large number of randomly generated models.  

\medskip

\begin{figure}[t]
  \centering
  \includegraphics[width=0.6\linewidth]{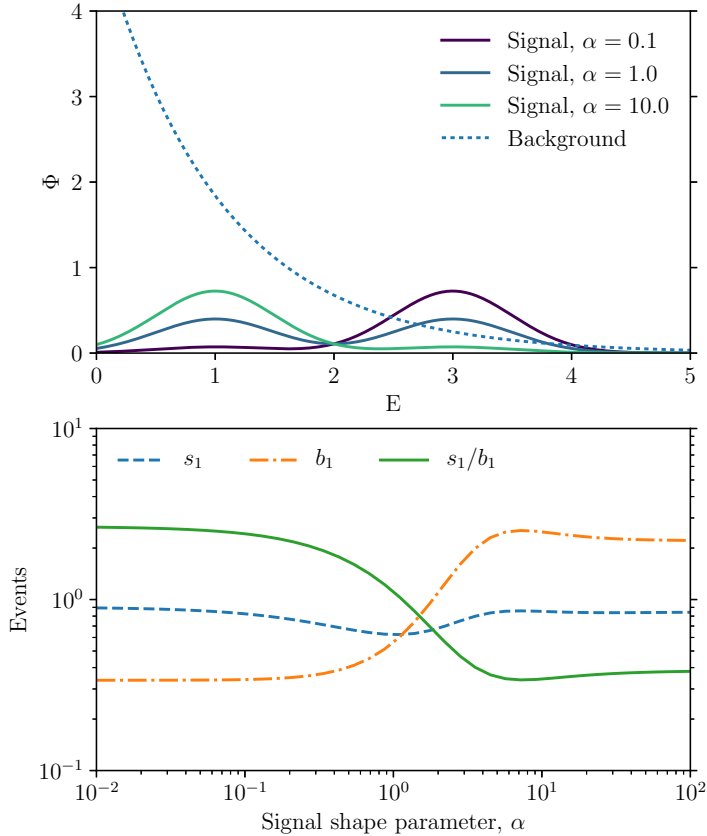}
  \caption{\textit{Upper panel:} Example spectra of Eq.~\eqref{eqn:Spectra} for various values of $\alpha$.  \textit{Lower panel:}  Illustration of equivalent signal counts $s_1$, background counts $b_1$, and SBR $s_1/b_1$ as function of the signal shape parameter $\alpha$.}
  \label{fig:spec}
\end{figure}

\paragraph*{Illustration.} Finally, we illustrate the definitions introduced in this subsection with a simple example.  Consider the following model
\begin{equation}
  \Phi(E) = \theta_1\left( \frac{\alpha}{1+\alpha}N_1(E) +\frac{1}{1+\alpha}N_3(E)\right) + \theta_2 e^{-E}\;,
\label{eqn:Spectra}
\end{equation}
where $\alpha$ is a shape parameter of the signal component, $N_1(E)$ and $N_3(E)$ are normal distributions with variance $0.5$ that are centered on the indexed values, and we set the signal normalization to $\theta_1=1$ and the background normalization to $\theta_2=8$.  Example spectra are shown in the upper panel of Fig.~\ref{fig:spec}, for various values of $\alpha$.  We neglect mixing between the components by assuming that the background is fixed via external constraints.

In the lower panel of Fig.~\ref{fig:spec} we show the corresponding equivalent signal and background counts of component $i=1$, as well as the corresponding SBR $s_1/b_1$.  For $\alpha\ll1$, the signal is dominated by the high energy peak, which is in a region of low background.  Indeed, we find approximately $b_1\simeq 0.33$ and $s_1\simeq 0.89$.  On the other hand, for $\alpha\gg1$, the signal is present at lower energies, where the background is larger.  Indeed, we find here $b_1\simeq 2.2$.

In the transition region, the equivalent number of signal events $s_1$, which otherwise remains close to one, drops somewhat.  This is expected, since at $\alpha\sim 1$ half of the signal is in a region of large backgrounds, whereas the other half is in a region with low backgrounds.  Since the low-background component dominates the signal-to-noise ratio in that case, it is also mostly this component that contributes to the equivalent signal number counts.  The effect becomes more pronounced if the background below $E<2$ is further increased, or above $E>2$ further reduced.

\section{Expected exclusion limits and discovery reach}
\label{sec:estimates}

In this section, we introduce prescriptions for deriving expected exclusion and discovery limits from the Fisher information matrix.  These prescriptions build on the equivalent signal and background counts that we defined above.  We refer to the corresponding prescriptions as Equivalent Counts Method (ECM).  For simplicity, we take the model component of interest to be $i=1$.  We put particular emphasis on the case of very small or a vanishing number of background events, and validate the approach for a few examples using MC simulations.  Caveats are discussed in the last subsection.

\subsection{Expected exclusion limits}

\paragraph*{Equivalent counts method.}
Projected approximate upper limits on the parameter $\theta_1$, assuming that the true value is $\theta_1=0$, can be derived from the Fisher information matrix by solving the equation
\begin{equation}
  s_1(\vect\theta^U) = Z(\alpha)\cdot\sqrt{s_1(\vect\theta^U) + b_1(\vect\theta^U)}\;,
  \label{eqn:ULsb}
\end{equation}
for $\theta_1^U$, while keeping the remaining $n-1$ parameters fixed to their respective values.  Here, we defined $\vect\theta^U = (\theta_1^U, \theta_2, \dots, \theta_n)^T$.  Furthermore, $s_i$ and $b_i$ refer to the equivalent signal and background counts that we introduced above in Sec.~\ref{sec:effsb}.  Finally, $Z(\alpha)$ is connected to the desired confidence level of the limit, $100(1-\alpha)\%$ CL, and $\alpha$ is the significance level.  It is derived from the inverse of the standard normal cumulative distribution distribution, denoted $F_{\mathcal{N}}$, as
\begin{equation}
  Z(\alpha) = F_{\mathcal{N}}^{-1}(1-\alpha)\;.
\end{equation}
In the case of, say, a $95\%$ CL upper limits, we have $\alpha=0.05$ and hence $Z=1.64$~\cite{Cowan:2010js}.  It is convenient to rewrite Eq.~\eqref{eqn:ULsb} as
\begin{equation}
  \theta_1^{U} =
  Z(\alpha)\cdot \sigma_1\left(\vect\theta^U\right)\;,
  \label{eqn:UL}
\end{equation}
which is in practice much easier to evaluate than Eq.~\eqref{eqn:ULsb}.

\medskip

\begin{figure}[t]
  \centering
  \includegraphics[width=0.6\linewidth]{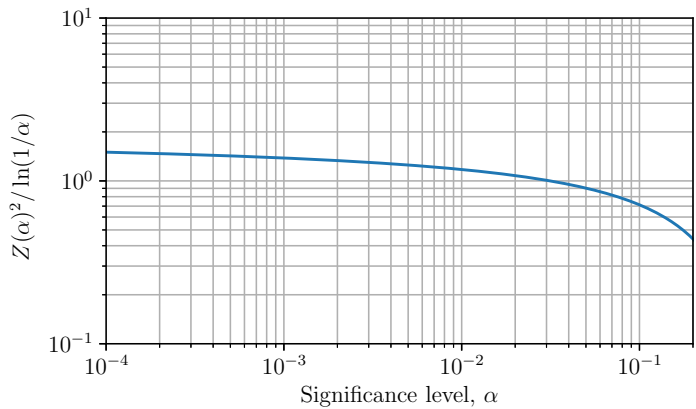}
  \caption{Ratio between expected exclusion limits from the ECM and the exact Neyman belt construction, assuming a single bin with vanishing expected background, as function of the significance level of the limit $\alpha$.  See Eq.~\eqref{eqn:ULapprox} and text for details.}
  \label{fig:fudge_factor}
\end{figure}

In the \emph{background-limited regime}, $\theta_1^U\Psi_1(E) \ll \Phi(E|\vect\theta)$, Eq.~\eqref{eqn:UL} implies that the exclusion limit satisfies
\begin{equation}
  \theta_1^{U} \simeq Z(\alpha)\cdot\sigma_1(\vect\theta)\;,
  \label{eqn:ULgaus}
\end{equation}
where $\vect\theta_0 = (0, \theta_2, \dots, \theta_n)^T$.  This is exactly what is expected if Gaussian background noise with variance $\sigma_1^2$ dominates.  In the \emph{signal-limited regime}, here defined as $\theta_1^U\Psi_1(E) \gg \sum_{i\geq2}\theta_i\Psi_i(E)$, Eq.~\eqref{eqn:UL} implies on the other hand
\begin{equation}
  \lambda_1(\theta_1^{U}) \simeq Z(\alpha)^2\;.
  \label{eqn:ULapprox}
\end{equation}
The upper limit is here independent of the signal model spectrum, $\Psi_1(E)$.  Eq.~\eqref{eqn:ULapprox} should be compared with $\lambda_1 = \ln 1/\alpha$,  which is the proper upper limit on the mean of a Poisson process when zero events are observed and the expected background is negligibly small.  We show in Fig.~\ref{fig:fudge_factor} that the fractional difference between $Z^2(\alpha)$ and $\ln(1/\alpha)$ is indeed small for typical significance level values used to set upper limits in the literature.  Namely, for $\alpha = 10^{-3}$--$10^{-1}$ the deviation is smaller than 38\%. 

We note that the general method to estimate expected upper limits in Eq.~\eqref{eqn:ULsb} works very well even in the presence of parameter mixing and background systematics as discussed in Sec.~\ref{sec:systematics}, as long as the associated changes of the background flux remain `sufficiently small' (say, below a few tens of percent) in the signal region.  A quantitative discussion can be found in subsection~\ref{sub:limitations} below.

\medskip

\paragraph*{Comparison with exact methods.}

The most general exact method for deriving upper limits with the correct coverage, and actually any sort of confidence intervals, is based on the Neyman Belt construction~\cite{Neyman:1937uhy} (an instructive overview can be found in Ref.~\cite{Feldman:1997qc}).  In practice, often the more specific MLR method is used to construct confidence intervals of any sort.  In the small-sample limit, MC simulations are required to establish the statistics of the MLR and construct intervals with the desired coverage (this is what we do here in all cases).  Details about the construction of confidence intervals using both methods can be found in Appendix~\ref{apx:UL}.

\begin{figure}[t]
  \centering
  \includegraphics[width=0.6\linewidth]{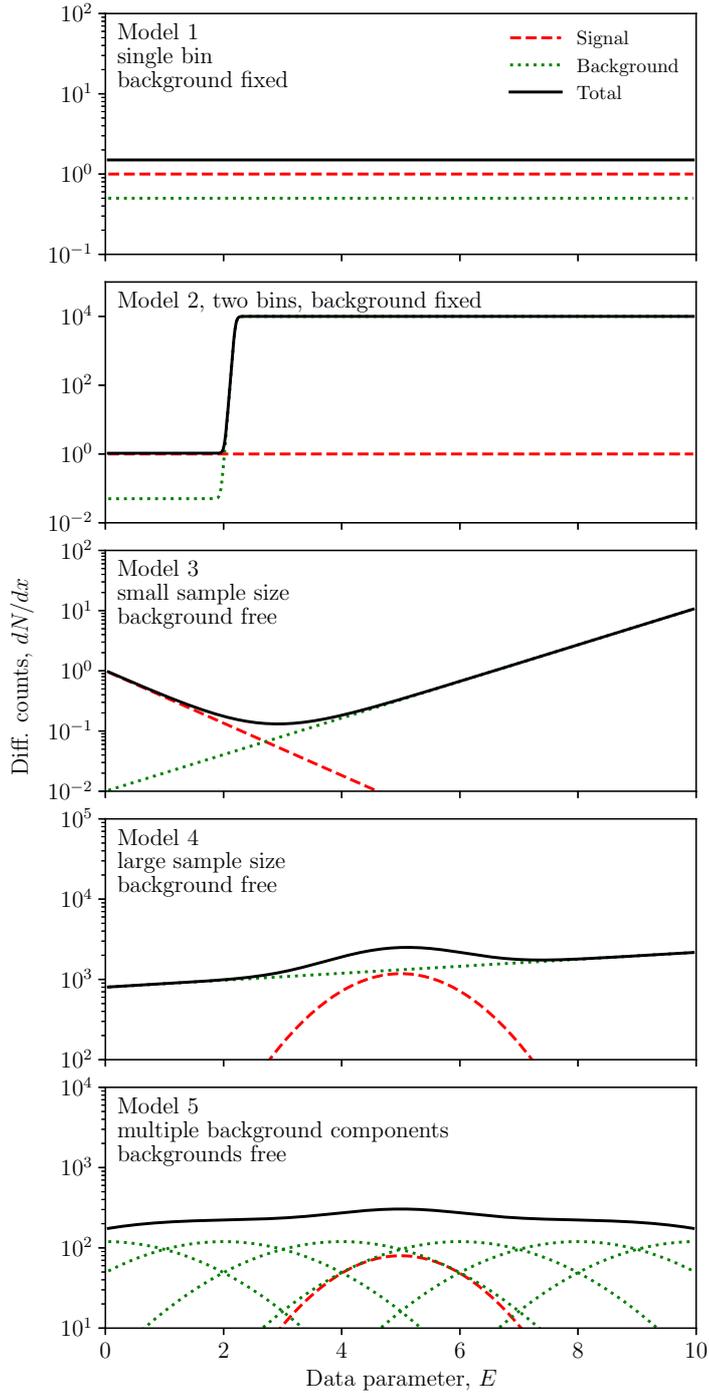}
  \caption{Five example models that we use to test the EC method against the full MLR method.  The signal component refers to $\Psi_1(E)$, the background component to $\Psi_2(E)$.  We also indicate whether background components are treated as fixed or free in the analysis.  Model 5 has six background components, $\Psi_{2-7}(E)$, instead of one.}
  \label{fig:Example_signals}
\end{figure}

\smallskip

As specific examples, we consider the signal and background functions shown in Fig.~\ref{fig:Example_signals}.  Model 1 is a simple single-bin example, whereas model 2 is a basic two-bin example with a strong difference in the expected background counts in both bins.  In both cases, we assume that the background normalization is known and fixed.  Model 3 provides an example in the small-sample regime, and model 4 in the large-sample regime.  The background normalization is here assumed to be determined by a fit to the data and hence free.  Model 5 is a scenario with multiple background components that are to some degree (but not completely) degenerate with the signal.

\begin{figure}[h]
  \centering
  \includegraphics[width=0.6\linewidth]{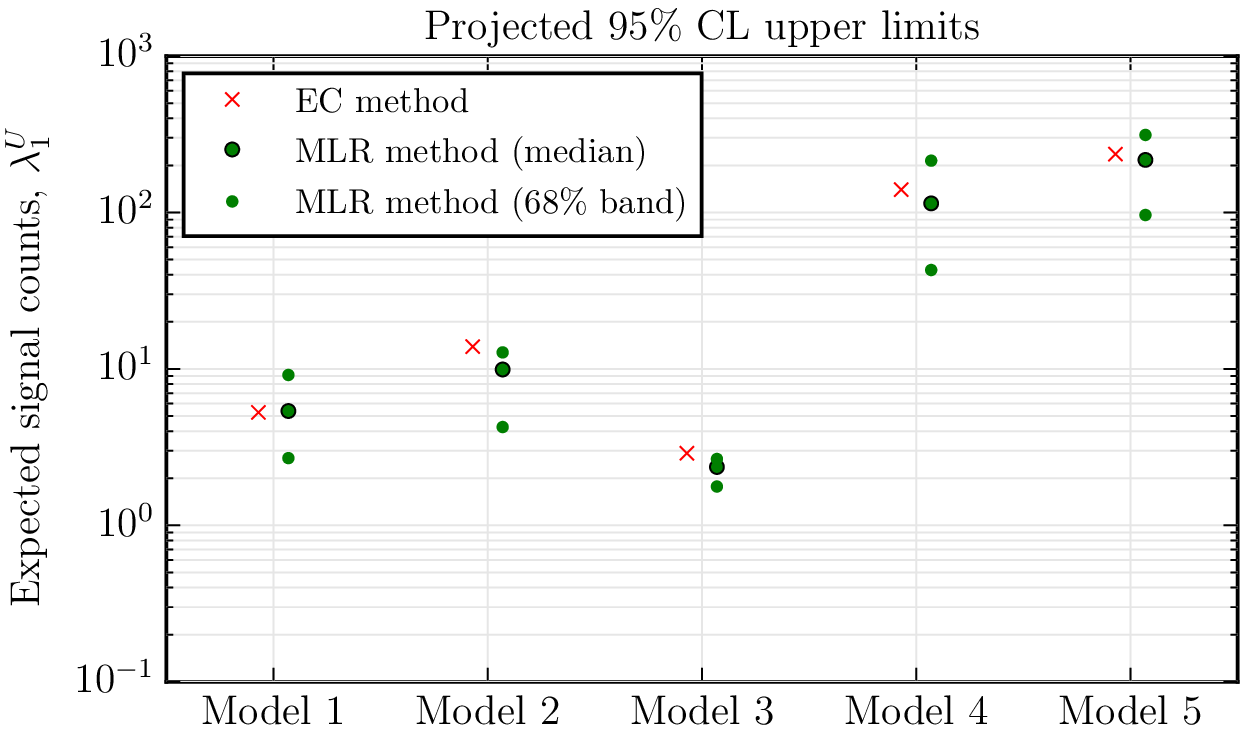}
  \caption{Expected exclusion limits for the five models in Fig.~\ref{fig:Example_signals}, derived using the ECM, Eq.~\eqref{eqn:UL}, compared to the median limit from the full coverage-corrected MLR method.  For comparison, we also show the 68\% containment regions of the MLR limits.  These regions contain, for multiple realizations of background-only data, $68\%$ of the corresponding upper limits.}
  \label{fig:MLims}
\end{figure}

In Fig.~\ref{fig:MLims}, we show the expected 95\% CL exclusion limits that we obtain from our EC method, Eq.~\eqref{eqn:UL}, for the five example models.  We compare these limits with the median limits obtained from the MLR method.  In three of the five cases (model 1, 4 \& 5) the agreement is remarkably good.  To emphasize this, we also indicate the extent of the $68\%$ containment band of the MLR limits, which contains $68\%$ of the upper limits when multiple realizations of the data are considered.  They are significantly wider than the difference between the limits from the EC and the MLR methods.

\begin{figure}[h]
  \centering
  \includegraphics[width=0.6\linewidth]{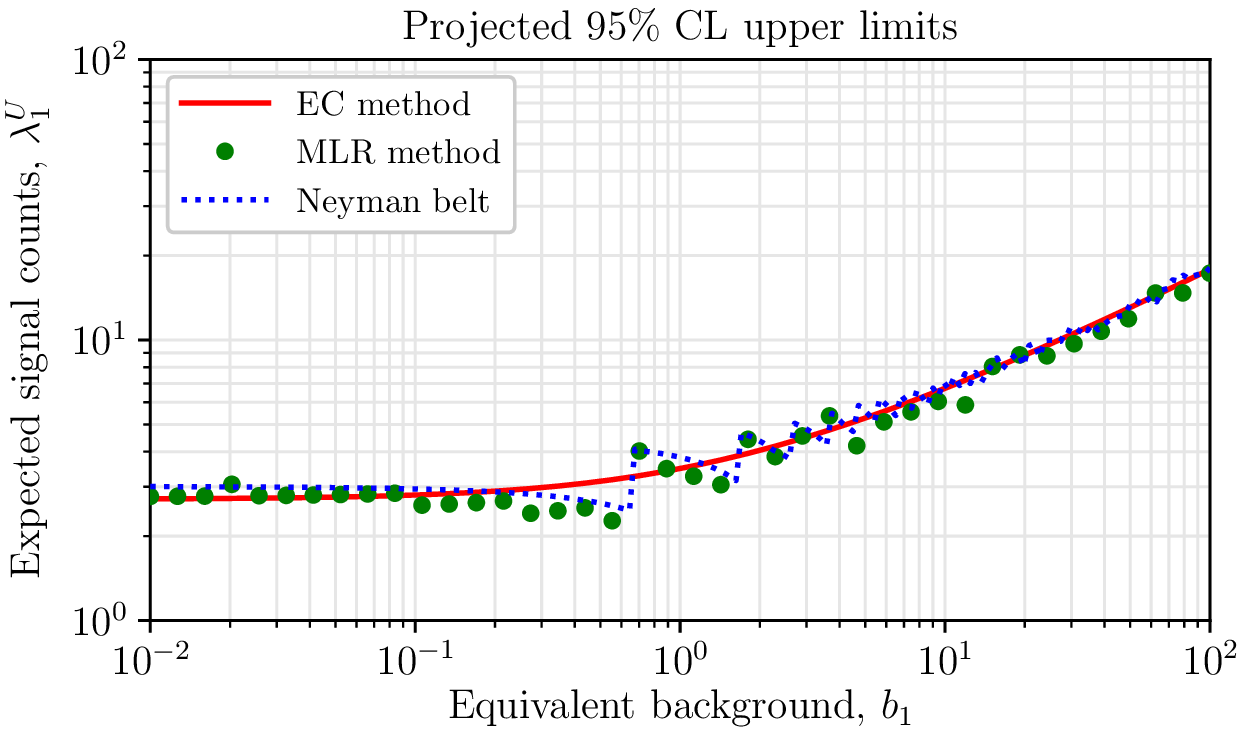}
  \caption{Various projected upper limits for model 1 from Fig.~\ref{fig:Example_signals}, as a function of the total equivalent background counts $b_1$:  Limits from the EC method (red solid line).  The median limit based on a full coverage-corrected MLR method (green dots).  The median limit from the Neyman Belt construction (blue dotted line).}
  \label{fig:UL1}
\end{figure}

It is instructive to discuss the single-bin model 1 in more detail.  In Fig.~\ref{fig:UL1}, we show the projected 95\% CL upper limits as derived from (1) the EC method, (2) the MLR method and (3) the full Neyman belt construction.   The upper limits are shown as function of the number of expected background counts.  In the large-sample limit, all methods give consistent results.  In the limit of vanishing background counts, the EC method yields slightly stronger projected limits than the Neyman belt construction (consistent with Fig.~\ref{fig:fudge_factor} and the above discussion).  In the intermediate regime, the Neyman belt construction shows a step-like structure, which is related to the discreteness of the Poisson likelihood.  This is not visible in the EC results, but is a small effect almost everywhere.  The MLR method yields results consistent with the Neyman belt construction.  Deviations are due to MC noise.

Finally, for models 2 and 3, we observe relevant differences between the methods.  The EC limits are here weaker (`more conservative') than the limits from the MLR method by up to $\sim40\%$.  We found that this is a rather general behaviour in the Poisson regime, which we observed for a large range of non-trivial scenarios.  However, as discussed above, these large deviations are not observed in the single-bin case model 1.

\begin{figure}[h]
  \centering
  \includegraphics[width=0.6\linewidth]{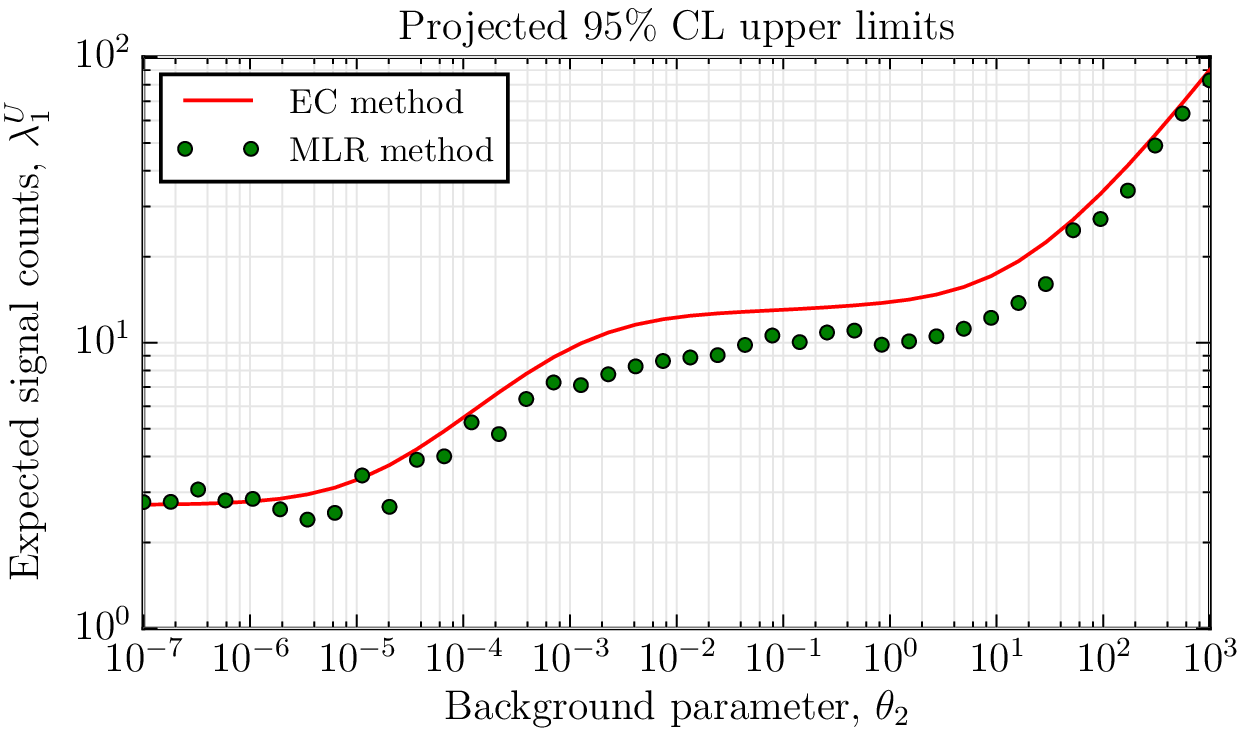}
  \caption{Projected upper limits for model 2 from Fig.~\ref{fig:Example_signals}, using the EC method and the coverage-corrected MLR method.    See text for a detailed discussion.}
  \label{fig:UL2}
\end{figure}

To shed further light on the difference between the single-bin and multi-bin scenarios in the Poisson regime, we show in Fig.~\ref{fig:UL2} for model 2 the expected exclusion limits derived from the EC and the MLR methods, as function of the normalization of the background component, $\theta_2$.  Three regimes can be clearly discriminated.  For $\theta_2\ll 10^{-5}$, the number of expected background counts over the entire range of $x$ is negligible. This case is equivalent to a simple single-bin scenario, where the EC and MLR methods agree well (see also Fig.~\ref{fig:UL1} for model 1).  On the other hand, for values of $\theta_2\gg 10^2$, we enter the Gaussian regime where the number of background counts is large over the entire range of $x$.  Again, EC and MLR results agree well.

However, in the intermediate range, $\theta_2\sim10^{-3}$--$10^{-1}$, we find a plateau where the MLR and EC methods yield different results.  In the plateau region, the number of background events is large at $2<x<10$, but negligible at $0<x<2$.  This effectively reduces the number of statistically relevant signal events by a factor of five, and should consequently weaken the expected exclusion limits by the same factor w.r.t.~the $\theta_2\ll 10^{-5}$ case.  This happens indeed for the limits derived using the EC method, but not for the limits from the MLR method.  We trace this behaviour back to the fact that in the plateau region, the events from $2<x<10$ introduce a noise in the otherwise Poissonian likelihood from $0<x<2$.  This effectively removes the discreetness of the Poisson distribution.  As a consequence, the MLR limits in the plateau region can cover exactly, while the MLR limits in the $\theta_2\ll 10^{-5}$ actually over-cover.

Scenarios like model 3 have a low- and a high-background regime, and hence behave similar to model 2 in the plateau region, leading to discrepancies between the MLR and the EC results (see Fig.~\ref{fig:MLims}).  We never found the effect to exceed $40\%$.

\subsection{Expected discovery reach}

\paragraph*{Equivalent counts method.}
The discovery limit for $\theta_1$, \ie~the value of $\theta_1$ that leads in $50\%$ of the cases to a rejection of the $\theta_1=0$ hypothesis with a significance level $\alpha$, can be approximately obtained by numerically solving the following equation for $\theta_1^D$:
\begin{equation}
  \label{eqn:mlr}
  \left(s_1(\vect\theta^D)+ b_1(\vect\theta^D)\right) \ln \left(\frac{ s_1(\vect\theta^D)+b_1(\vect\theta^D)}{b_1(\vect\theta^D)}\right)\\
  - s_1(\vect\theta^D) = \frac{Z(\alpha)^2}{2}\;.
\end{equation}
Here, we use the notation $\vect\theta^D= (\theta_1^D, \theta_2, \dots, \theta_n)^T$, and $s_1$ and $b_1$ refer to the equivalent signal and background counts discussed at the end of Sec.~\ref{sec:effsb}.  Heuristically, Eq.~\eqref{eqn:mlr} is motivated by the analytic structure of profile likelihood ratios, see for instance discussion in Ref.~\cite{Cowan:2010js}.  However, its main motivation comes from the fact that it leads to approximately correct results both in the signal- and background- limited regimes.

\medskip

It is useful to consider limiting cases.  In the \emph{background-limited case}, $b_1(\vect\theta^D) \gg s_1(\vect\theta^D)$, Eq.~\eqref{eqn:mlr} implies that
\begin{equation}
  \theta_1^{D} \simeq Z(\alpha)\cdot\sigma_1(\vect\theta)\;,
  \label{eqn:Gauss_approx}
\end{equation}
with $\vect\theta_0 = (0, \theta_2, \dots, \theta_n)^T$.  This is exactly what we expect for Gaussian background noise with variance $\sigma_1^2$.  On the other hand, in the \emph{signal-limited case}, $b_1(\vect\theta^D) \ll s_1(\vect\theta^D)$, one can show that the solution to Eq.~\eqref{eqn:mlr} satisfies the following equation (details can be found in Appendix~\ref{apx:gammaD}):
\begin{equation}
  \frac{b_1(\vect\theta^D)^{s_1(\vect\theta^D)}}
  {\Gamma\left(s_1 (\vect\theta^D)+1\right)}
  = \alpha \cdot \sqrt{\frac{Z^2(\alpha)}{s_1(\vect\theta^D)}}\;.
  \label{eqn:poisAprox}
\end{equation}
Here, $\Gamma(\cdot)$ is the gamma function.  This equation is very similar to the exact expression derived from the Poisson distribution using Asimov data~\cite{Cowan:2010js} in the low-background limit.  To see this, note that, given some background $b_1 \ll 1$ and zero signal, the probability to detect $\lambda$ or more photons is approximately $(b_1)^\lambda/\Gamma(\lambda+1)$.  If $\lambda$ were the discovery reach corresponding to significance level $\alpha$, this expression should equal $\alpha$.  The difference between this exact and the above approximate expressions is hence the square-root on the right-hand side of Eq.~\eqref{eqn:poisAprox}.  In practice, this turns out to be a small effect, as we will see below.

\medskip

\paragraph*{Comparison with exact methods.}

\begin{figure}[h]
  \centering
  \includegraphics[width=0.6\linewidth]{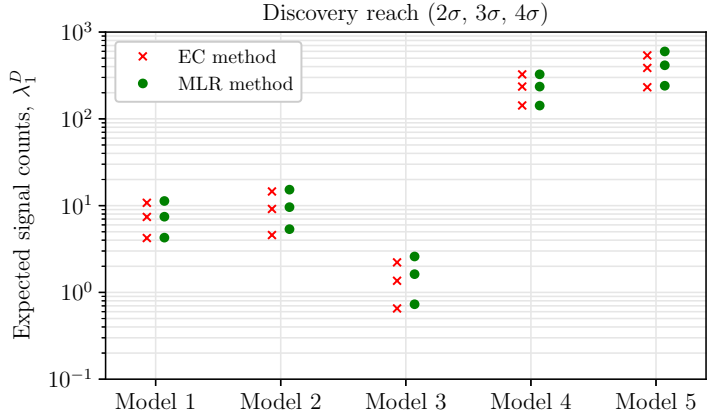}
  \caption{Expected discovery limits derived from the EC method, Eq.~\eqref{eqn:mlr}, compared wit full MLR results, for the five models in Fig.~\ref{fig:Example_signals}.  Groups of three symbols show from bottom to top $2\sigma$, $3\sigma$ and $4\sigma$ discovery limits (meaning that 50\% of the measurements would lead to a detection with at least the indicated significance).  Both methods yield consistent results, see detailed discussion in the text.}
  \label{fig:MDT}
\end{figure}

In Fig.~\ref{fig:MDT} we compare the expected discovery limits derived using the EC method, Eq.~\eqref{eqn:mlr}, with the ones from the full MLR method, for the five example models in Fig.~\ref{fig:Example_signals} (details can be found in Appendix~\ref{apx:UL}).  We find that for all cases the EC results remain very close to the MLR results. Deviations are usually much less than $1\sigma$, and largest for model 3 where the equivalent background is lowest.  This good agreement is quite remarkable, given that some of the models are deeply in the Poisson regime.

\begin{figure}[h]
  \centering
  \includegraphics[width=0.6\linewidth]{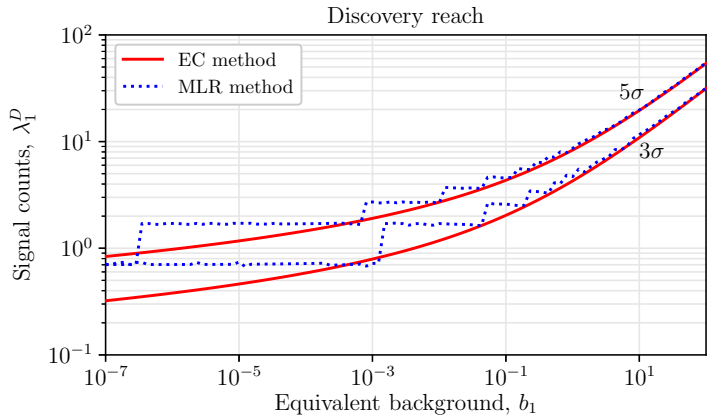}
  \caption{Comparison of $3\sigma$ (bottom) and $5\sigma$ (top) expected discovery limits based on the EC method, Eq.~\eqref{eqn:mlr}, and on the MLR method. The discreteness of the Poisson distribution becomes apparent for $b_1\lesssim1$, where the pure EC methods over-predicts the sensitivity by the indicated amount.  See text for a detailed discussion.}
  \label{fig:det1}
\end{figure}

To further investigate the limitations of our EC method in the low-background regime, we compare in Fig.~\ref{fig:det1} for model 1 the $3\sigma$ and $5\sigma$ expected discovery limits derived from the EC and the MLR methods.  The approximate discovery limits resemble well the exact results down to equivalent backgrounds of one.  For smaller equivalent backgrounds, the discreteness of the Poisson distribution starts to dominate the MLR method, which is not seen in the EC results.  However, depending on the significance level, the agreement remains reasonably good even down to and below $b_1\sim 10^{-3}$.  We note that, in very extreme cases, the EC method could lead to expected discovery limits that are much smaller than one.  This can be prevented by the additional ad hoc requirement that the equivalent number of signal counts, $s_1(\vect\theta)$, should be at least one.

\subsection{Limitations of the Fisher approach}
\label{sub:limitations}

There are two important situation where the prescriptions for deriving expected exclusion limits, Eq.~\eqref{eqn:UL}, and expected discovery limits, Eq.~\eqref{eqn:mlr}, will break down and give potentially wrong results.\footnote{This is the case for the additive component models considered in the present work.  For more general Poisson problems, \fex~with shape uncertainties, the number of potential problems is larger.} Here we address both in detail.

\medskip

\paragraph*{Skewness of Poisson likelihood.}  The Fisher information matrix as defined in Eq.~\eqref{eqn:I} encodes complete knowledge about the likelihood function, provided that the likelihood behaves like a multivariate normal distribution.  This is equivalent to requiring that higher order derivatives in the expansion of the log-likelihood are negligible.  Often, higher derivative terms will only cause deformations to Gaussian contours, but in the extreme case in which parameters are infinitely degenerate, the Fisher formalism will provide not only quantitatively but also qualitatively wrong results~\cite{Wolz:2012sr}.  Non-Gaussianity is a common challenge in cosmology~\cite{SellentinQuartinAmendola2014}, but also relevant for the simple additive component model with the Poisson likelihood considered here.

As we saw above, using our EC method, one can obtain reasonable expected exclusion and discovery limits even in the deep Poisson regime, despite the fact that the likelihood functions are clearly non-Gaussian in that case.  This is partially due to some lucky numerical coincides that happen to make our proposed prescriptions reasonably accurate. However, these mechanisms only apply to the signal-component of interest.  The likelihood functions corresponding to the background (any non-signal) components should obey the usual Gaussianity constraints to ensure that the EC method can be applied.

In Appendix~\ref{apx:Gregime}, we study the behaviour of the Poisson likelihood function up to third order in the model parameters.  Assuming that no strong degeneracies exist between parameters, one can derive a simple condition on the equivalent number of background events that should hold for all background components.  It reads
\begin{equation}
  b_i \gtrsim \frac{4t}{9f^2}\;,
  \label{eqn:b_cond}
\end{equation}
where $b_i$ is the number of \emph{equivalent background} counts of component $i$, $t$ the value of $\Delta(2\ln\mathcal{L})$ at the boundary of the confidence region of interest, and $f$ the tolerable fractional uncertainty of $t$ at that boundary.  If, for instance, we want to have $2\sigma$ intervals (for one dimension this corresponds to $t=4$) with a fractional significance error of less than $20\%$ (this corresponds to $f\simeq 40\%$, since $t$ depends quadratically on the significance in standard deviations), this implies that the equivalent background for the components $i$ should exceed $b_i \gtrsim 11$.  In cases where parameter degeneracies might play a role, we recommend to use the full expressions provided in Appendix~\ref{apx:Gregime} instead.

\medskip

\paragraph*{Parameter degeneracies.}

When using the Fisher formalism, implicitly assumes that model parameters are unbound and only constrained by the likelihood function.  This means that there is nothing that prevents part of the parameters $\theta_i$ to become negative.  Consider as simple example a signal, $\Psi_i(E)$, that is exactly degenerate with a background component $k\neq i$, $\Psi_k(E) \propto \Psi_i(E)$.  In that case, the Poisson part of the Fisher information matrix becomes singular, and the error of the signal component, $\sigma_i$, diverges.  An arbitrarily large signal $i$ can be compensated by an equally large negative background component $k$.  Since this behaviour is usually unphysical (unless for instance absorption effects are part of the model), it must be prevented when performing Fisher forecasting.

A sufficient condition to exclude the problems with negative components reads
\begin{equation}
  s \cdot \sigma_k(\vect\theta) <\theta_k \quad \text{for all} \quad k\neq i\;,
  \label{eqn:pdc}
\end{equation}
where $s$ is the significance of interest in standard deviations.  If errors are sufficiently small, the parameter $\theta_k$ will not cross zero.  As will be discussed below in section~\ref{sec:systematics}, if the projected data alone is not sufficient to break the degeneracy between different background components or the background and the signal, it is possible (and necessary) to include additional constraints on parameters such that Eq.~\eqref{eqn:pdc} is fulfilled.  If this is not possible, the EC method cannot be directly applied.

\section{Modeling of instrumental and background systematics}
\label{sec:systematics}

We give here a few instructive examples of how to model background uncertainties within the Fisher information framework.

\subsection{Basic parameter systematics}
\label{sec:syst_basics}

In many cases of practical importance, additional information about nuisance parameters is available, which must be included in the sensitivity projections to obtain realistic results.  Within a Bayesian approach, these additional constraints would be included as priors on the nuisance parameters.  Within the Frequentist treatment, which we focus on here, a common approach is to include additional parameter constraints as effective likelihoods, as described in the following.  These additional constraints can, for instance, come from `sideband measurements' in signal-free regions of the data space.

If we assume that the constraints on parameter $\theta_i$ are Gaussian, with standard deviation $\xi_i$, the associated combined likelihood function can be written as
\begin{equation}
  \mathcal{L}(\mathcal{D}|\vect\theta) = \mathcal{L}(\mathcal{D}|\vect\theta)_\text{pois}\times
  \prod_i \mathcal{N}(\theta_i^A|\mu=\theta_i, \sigma^2=\xi_i^2)\;.
  \label{eqn:LLL}
\end{equation}
Here, $\mathcal{N}$ refers to the PDF of a normal distribution with variance $\xi_i^2$ and Asimov value $\theta_i^A$.  It accounts for potential sideband measurements and similar external constraints.  In this spirit, $\theta_i^A$ is taken to equal the mean value, $\theta_i^A=\theta_i$, when the average $\langle \cdot \rangle_{\mathcal{D}(\vect\theta)}$ is applied in Eq.~\eqref{eqn:I}. 

The resulting total Fisher information matrix can be split in a Poisson and a systematics part, $\mathcal{I}_{ij} = \mathcal{I}_{ij}^\text{pois} + \mathcal{I}_{ij}^\text{syst}$, where the latter is here diagonal and given by
\begin{equation}
  \mathcal{I}_{ij}^\text{syst}= \delta_{ij}\frac{1}{\xi_i^2}\;.
  \label{eqn:Isyst}
\end{equation}
The generalization to correlated systematic errors reads $\mathcal{I}_{ij}^\text{syst}= (\Sigma_\text{syst}^{-1})_{ij}$.  We will show in a few examples how this is used in practice.

\subsection{Example 1: Background systematics degenerate with the signal}

We start with a simple example where we assume that some component of the background systematic is perfectly degenerate with the signal.  More specifically, we consider a three-component model, where $\Psi_1(E)$ is the signal, $\Psi_2(E)$ the nominal background, and the component $\Psi_3(E) = \Psi_1(E)$ accounts for small positive or negative perturbations of this background.  Note that this implies that $\mathcal{I}^\text{pois}_{1i} = \mathcal{I}^\text{pois}_{3i}$ for $i=1,2,3$, which means that the Poisson part of the Fisher matrix is singular.  We set the background normalization to $\theta_2=1$, and the mean background perturbation to zero, $\theta_3=0$.  For the systematics part of the Fisher matrix, Eq.~\eqref{eqn:Isyst}, we assume that the background perturbation $\theta_3$ is externally constrained with a variance of $\xi^2_3>0$, and the background normalization $\theta_2$ and the signal normalization are unconstrained, $\xi_1^2,\xi_2^2\to\infty$.

In that case, one can show, by calculating the profiled Fisher information for component $i=1$, that the variance of the signal component is given by (details can be found in Appendix~\ref{apx:syst_errors})
\begin{equation}
  \sigma_1^2(\vect\theta) = (\sigma_1^\text{pois})^2(\vect\theta)+\xi_3^2\;.
  \label{eqn:sigma_stat_syst}
\end{equation}
This means that, as one might have expected, statistical and systematic errors are added in quadrature.  We remark that constraints on the signal parameter have the opposite effect, $\sigma_1^{-2}(\vect\theta) = (\sigma_1^\text{pois})^{-2}(\vect\theta)+\xi_1^{-2}$, and \textit{decrease} the overall variance.

\begin{figure}[t]
  \centering
  \includegraphics[width=0.6\linewidth]{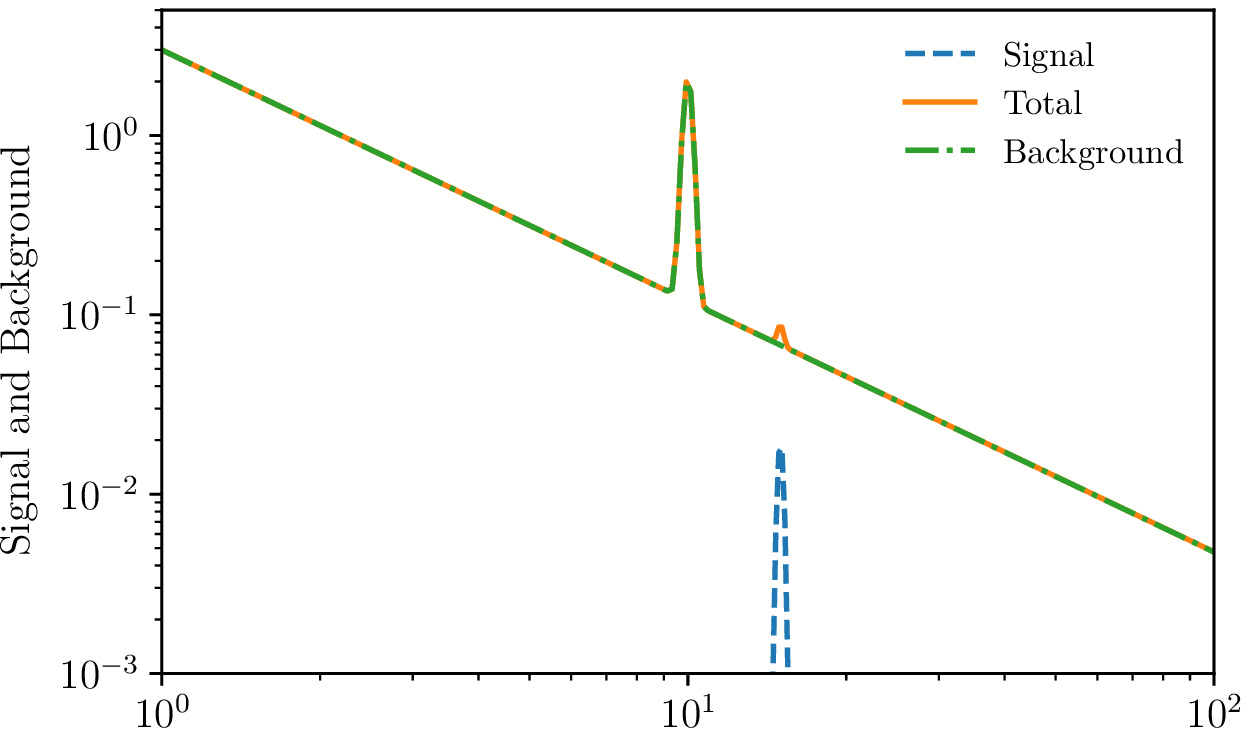}
    \includegraphics[width=0.6\linewidth]{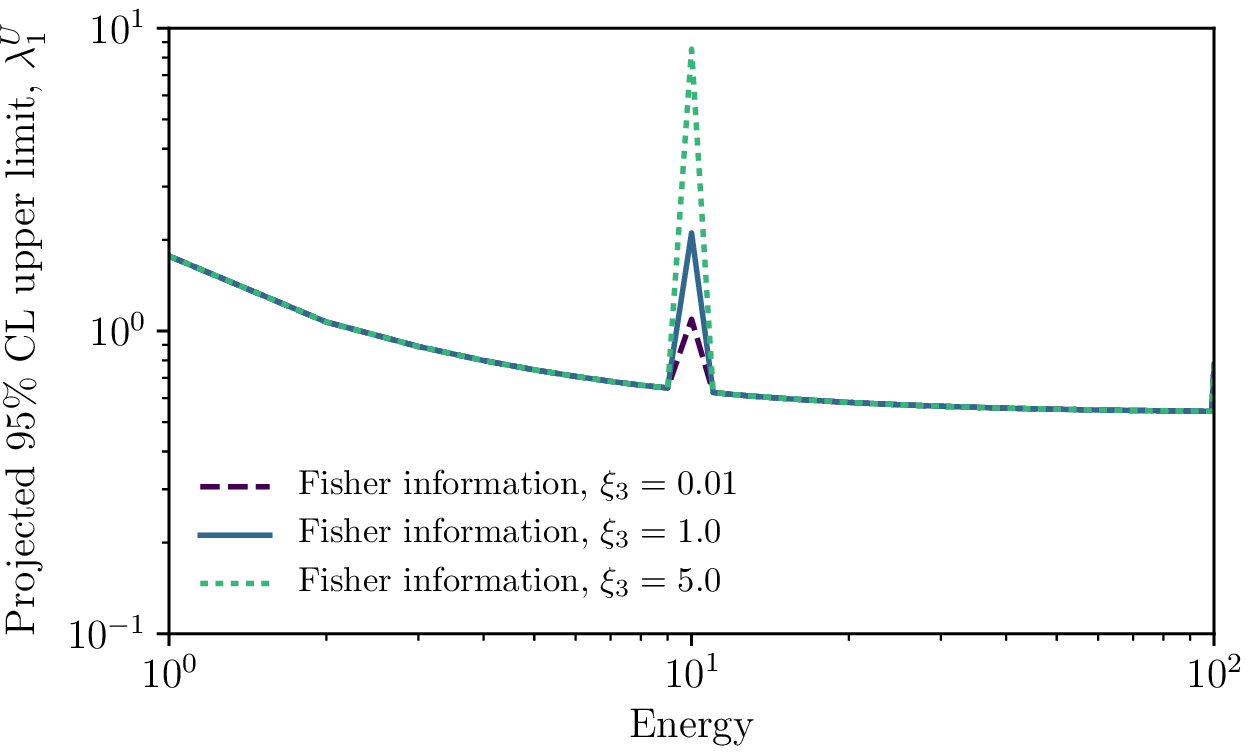}
    \caption{\emph{Upper panel:} Example of a fixed power law background with one instrumental lines at $E=10$.  We show an example of a signal line with identical width to the instrumental lines (here at $E=15$). \emph{Lower panel:} We calculate expected exclusion limit using the EC method, as a function of the signal position. The limits becomes more constraining at higher energies, due to the increased SBR. However, there is loss of constraining power when the signal is degenerate with the instrumental line, depending on the uncertainty of the instrumental line, $\xi_3$.}
  \label{fig:signal_sec_IV}
\end{figure}

\medskip

To illustrate Eq.~\eqref{eqn:sigma_stat_syst}, we consider expected exclusion limits on line-like signals on top of a power-law background plus an `instrumental line'.  We treat the power-law background as fixed, and assume that the instrumental line has identical width to the signal line. However, the normalization of the instrumental line is constrained by $\xi_3 = 1.0$ . The fluxes of the background and signal components are illustrated in the upper panel of Fig.~\ref{fig:signal_sec_IV}. Specifically we consider a Gaussian located at $E=10$, with width $\sigma^2_E = 0.2$. The power law has a slope of $1.4$ and normalisation of three, $\theta_2=3.0$. The normalisation of the instrumental line is kept fixed with $\theta_3=1.0$.

The projected limits are shown in the lower panel of Fig.~\ref{fig:signal_sec_IV}.  When the signal becomes degenerate the with instrumental lines, the projected limits weaken, according to Eq.~\eqref{eqn:sigma_stat_syst}. We also recover the expected behaviour that when the normalization of the instrumental line is unconstrained (see Sec.~\ref{sub:limitations}) , the projected limit calculated by the Fisher information at the position of the instrumental line diverges.

\subsection{Example 2: Background systematics described by correlation function}
\label{sec:corr_syst}

A more general approach towards the modeling of background systematics is to write the total flux as
\begin{equation}
  \Phi(E) = \theta_1\Psi_1(E) + (1+\delta(E)) \Psi_2(E)\;,
  \label{eqn:flux_gaussian_errors}
\end{equation}
where again $\Psi_1$ denotes the signal, $\Psi_2$ the nominal background (we fix $\theta_2=1$ throughout), and $\delta(E)$ parameterizes small deviations from the background model.  In general, systematic uncertainties in the background will be correlated as a function of energy, and one can define the covariance function
\begin{equation}
  \langle \delta(E) \delta(E') \rangle = \Sigma_\delta(E,E')\;,
\end{equation}
and we assume $\langle \delta(E)\rangle=0$.  More specifically, $\delta(E)$ can
be thought of as \textit{Gaussian random field} with mean zero, whose behaviour
is completely determined by the covariance function. It incorporates
information both about the variance and correlation of the systematic.

For any practical calculations, one needs to discretize the field $\delta(E)$.  One simple way of doing that is to write $\delta(E)$ as a step function,
\begin{equation}
  \delta(E) = \sum_{i=1}^{N} \xi_i \chi_{\Delta E_i}(E)\;,
  \label{eqn:Delta}
\end{equation}
with the selector function
\begin{equation}
  \chi_{\Delta E_i} = \left\{
  \begin{array}{cc}
    1 & \text{if}\; E \in \Delta E_i  \\
    0 & \text{if}\; E \notin \Delta E_i
  \end{array}
\right.
  \;,
\end{equation}
where the (very small) energy bins $\Delta E_i$ cover the entire energy range of interest, and $E_i$ will denote the corresponding mean of each energy bin.  Furthermore, $\xi_i$ are constrained by a multivariate normal distribution, with a covariance matrix defined by $\Sigma_\delta(E_i, E_j)$.  The free parameters in the present example are then $\vect\theta = (\theta_1, \xi_1, \dots, \xi_N)$.

One can now show that the resulting profiled Fisher information for the signal (assuming a sufficiently fine binning which captures all relevant structure of $\Psi_1(E)$), is given by (details can be found in Appendix~\ref{apx:syst_errors})
\begin{equation}
  \mathcal{\widetilde I}_{11} = \sum_{ij}
  \frac{\Psi_1}{\Psi_2}(E_i)
  D^{-1}_{ij}
  \frac{\Psi_1}{\Psi_2}(E_j)\;,
  \label{eqn:I_corr_syst}
\end{equation}
where we defined the combined covariance matrix
\begin{equation}
 \label{eqn:D_cov}
  D_{ij} \equiv \frac{\delta_{ij}\Phi(E_i)}{\Delta E_i [\Psi_2(E_i)]^2} + \Sigma_\delta(E_i, E_j)\;,
\end{equation}
which includes both the effects of Poisson noise and background systematics. This expression becomes more simple in the regime where $\Psi_1 \ll \Psi_2$ since then $\Phi(E) = \Psi_2(E)$. In the limit of no systematics, $\Sigma_\delta \to 0$, we recover the standard expression for the Fisher information of the signal component, Eq.~\eqref{eqn:Ipsi}.  On the other hand, in the large sample limit, where $\Psi_2$ becomes large and hence the first term in Eq.~\eqref{eqn:D_cov} small, only the covariance matrix $\Sigma_\delta$ matters and determines the limiting accuracy at which $\Psi_1$ can be measured.   Note that results are independent of the bin size as long as it is sufficiently small to fully resolve the discriminating aspects of the different model components.

This approach of estimating the effect of background uncertainties on expected experimental sensitivities has been already used by some of the present authors in Ref.~\cite{Bartels2017}.  We will provide another example below in Sec.~\ref{sec:optimization}, in context of the Fisher information flux.

\section{Strategy optimization}
\label{sec:optimization}

Experimental design, or the planning of astronomical observational campaigns, often make use of the SNR of some signals of interest (for the simple additive component models, Eq.~\eqref{eqn:add_model} that we discussed above, this corresponds to $\propto\Psi_i(E)/\sqrt{\Phi(E)}$).  One of the goals is to maximize the exposure of energy and/or spatial regions that provide the largest SNR for some component $i$, which then leads to the tightest constraints on the model parameter $\theta_i$.

We will show here that the above SNR is the simplest realization of the Fisher \emph{information flux}, which we newly introduce here.  However, the latter is much more general and can also naturally account for the non-local and saturation effects of background and instrumental systematics.  We will illustrate this in an example that makes use of the treatment of correlated background systematics that we discussed in the previous section.

\subsection{Fisher information flux}

In this subsection we will look at the model spectrum $\Phi$ as function of the sky coordinate, $\Phi(\Omega|\vect\theta)$, such that $I_i(\Omega)$ is the intensity of the signal or background component,\footnote{To keep the notation simple, we ignore here the effects of the instrument point-spread function or energy dispersion.  They can be added straightforwardly.} and $\epo(\Omega)$ is the exposure of the instrument towards $\Omega$, see Eq.~\eqref{eqn:def_epo}.

With this, we can define the differential \textit{Fisher information flux} that corresponds to parameter pair $(i,j)$ as functional derivative w.r.t.~the exposure at position $\Omega$.  It is given by
\begin{equation}
  \frac{d\mathcal{F}_{ij}}{d\Omega}(\vect\theta, \mathcal{E}) \equiv \frac{\delta\mathcal{I}_{ij}(\vect\theta, \mathcal{E})}{\delta\epo(\Omega)}\;,
  \label{eqn:Iflux}
\end{equation}
where we made explicit that the Fisher information is in general a function of the exposure map $\mathcal{E}(\Omega)$.  If we consider the Poisson part of the Fisher information alone, we find
\begin{equation}
  \frac{\delta\mathcal{I}_{ij}^\text{pois}(\vect\theta, \mathcal{E})}{\delta\epo(\Omega)} =
  \frac{1}{\theta_i\theta_j}
  \frac{I_i(\Omega) I_j(\Omega)}
  {I(\Omega|\vect\theta)}\;.
  \label{eqn:IfluxP}
\end{equation}
The diagonal part of the Fisher information flux corresponds here to the SNR of component $i$, and the non-diagonal parts provide information about the degeneracy of the components pairs $(i,j)$.  

We emphasize that the right-hand side of Eq.~\eqref{eqn:IfluxP} does \emph{not} depend on the exposure anymore, and is hence constant during the course of the measurement.  Similarly, external constraints like in Eq.~\eqref{eqn:Isyst} do not depend on the exposure.  In these cases, the full Fisher information flux in Eq.~\eqref{eqn:Iflux} would equal the Fisher information flux of the Poisson likelihood, Eq.~\eqref{eqn:IfluxP}.  This simple situation changes drastically when considering the \emph{effective information flux} for a subset of the model parameters, as we will see in the next subsection.

\medskip

The information gain about the parameter pair $(i,j)$ that is obtained by increasing the exposure towards direction $\Omega$ by the infinitesimal amount $\delta\mathcal{E}(\Omega)$ is given by
\begin{equation}
  \delta\mathcal{I}_{ij}(\vect\theta, \mathcal{E}) =
  \int d\Omega\,
  \delta\mathcal{E}(\Omega)
  \frac{d\mathcal{F}_{ij}}{d\Omega}(\vect\theta, \mathcal{E})\;.
\end{equation}
As a simple application, let us assume that the change in exposure per time is given by
\begin{equation}
  \frac{d\mathcal{E}(\Omega)}{dt} = \aeff \frac{d\tobs(\Omega)}{dt}\;,
\end{equation}
where we factored out the effective area, $\aeff$, and $\tobs(\Omega)$ is the accumulated observation time in direction $\Omega$. Then, the
information gain per unit time is given by
\begin{equation}
  \frac{d\mathcal{I}_{ij}(\vect\theta, \mathcal{E})}{dt} =
  \aeff \int d\Omega\,
  \frac{d\tobs(\Omega)}{dt}
  \frac{d\mathcal{F}_{ij}}{d\Omega}(\vect\theta, \mathcal{E})\;.
\end{equation}
Integrating this over time would again give the total information obtained in the observation.  Note that these equations remain valid also for the effective information flux that we discuss in the next subsection.

\subsection{Effective information flux}
\label{sub:marg_information_flux}

In order to quantify information gain about PoIs in presence of background and other uncertainties, the above concept of Fisher information flux needs to be extended to the profiled Fisher information that we introduced in Eq.~\eqref{eqn:Imarg}.  This can be done straightforwardly by applying the functional derivative with respect to $\epo(\Omega)$, Eq.~\eqref{eqn:Iflux}, to the profiled Fisher information.

The full expression for the \emph{effective information flux} for the PoI $(\theta_1, \dots, \theta_k)$ reads
\begin{equation}
  \frac{d\mathcal{\widetilde F}_A}{d\Omega} =
  \frac{d\mathcal{F}_A}{d\Omega}
  - \frac{d\mathcal{F}_C^T}{d\Omega}\, \mathcal{I}_B^{-1}\, \mathcal{I}_C 
  +\mathcal{I}_C^T\, \mathcal{I}_B^{-1}\, \frac{d\mathcal{F}_B}{d\Omega}\, \mathcal{I}_B^{-1}\, \mathcal{I}_C
  - \mathcal{I}_C^T\, \mathcal{I}_B^{-1}\, \frac{d\mathcal{F}_C}{d\Omega}\;,
  \label{eqn:mF}
\end{equation}
where as above $A$ refers to the $k\times k$ part of the Fisher information matrix that corresponds to the PoI, $B$ refers to the $(n-k)\times(n-k)$ part for the nuisance parameters, and $C$ to the mixing between nuisance parameters and the PoI.  This expression appears lengthy, but straightforward to evaluate analytically or numerically if the Fisher information matrix and Fisher information flux are already known.

\medskip

The effective information flux has a number of surprising and useful properties.  First, it is in general not constant in time (in contrast to the plain information flux in Eq.~\eqref{eqn:Iflux}).  In the examples considered here, this is due to \emph{saturation effects}, which are related to the observation reaching the systematic limited regime.   Second, it is \emph{non-local}, in the sense that it for instance depends on the past (non-)observation of sidebands that could help to characterize the backgrounds in the signal region.  Technically, the non-locality due  to the fact that the full Fisher information matrix appears in Eq.~\eqref{eqn:mF}, which includes integrals of the signal and background intensities over $\Omega$.  We will illustrate these two aspects in the following final example.

\medskip

\paragraph*{Saturation effects.}
We demonstrate the saturation effects of the effective information flux with a non-trivial examples from Sec.~\ref{sec:corr_syst}.  There, we discussed how to treat correlated background systematics around a fixed background in the Fisher formalism, see Eq.~\eqref{eqn:flux_gaussian_errors}.

For definiteness,
we consider a signal component that consists of two Gaussian peaks,
\begin{equation}
  I_1(E)= 0.05\times\mathcal{N}(E|\mu=2, \sigma^2 = 0.01)\\ + \mathcal{N}(E|\mu=6, \sigma^2=4)\;,
\end{equation}
one of which is narrow and the other wide.  The background is taken to be flat, $I_2(E) = 1$.  The corresponding model count spectra $\Psi_{1,2}(E)$ are obtained by multiplication with the exposure $\mathcal{E}(E)$, as in Eq.~\eqref{eqn:def_epo}.   Lastly, we define the covariance function of the background deviations as,
\begin{equation}
  \Sigma_{\delta}(E,E') = 0.01 \times \mathcal{N}(E|\mu=E', \sigma^2 =1) + 0.01 \times \mathcal{N}(E|\mu=E', \sigma^2 =2)\;.
\end{equation}
For simplicity, we focus on the case $\theta_1=1$. A sum of two Gaussians was chosen to demonstrate the techniques ability to account for multiple correlation lengths.

The profiled Fisher information for the signal component $\Psi_1(E)$ is given by Eq.~\eqref{eqn:I_corr_syst} above.  The corresponding effective differential information flux at energy $E_k$ can be obtained by differentiating this expression w.r.t.~exposure in energy bin $\Delta E_k$ (remember that we have bins in energy for practical purposes).  The resulting \emph{effective information flux} of the signal reads
\begin{equation}
  \frac{d\mathcal{\widetilde F}_{11}}{dE}(E_k) = \\
  \sum_{ij}
  \frac{I_1}{I_2}(E_i)
  D^{-1}_{ik}
  \frac{I(E_k)}{\Delta E_k^2 \mathcal{E}(E_k)^2I^2_2(E_k)}
  D^{-1}_{kj}
  \frac{I_1}{I_2}(E_j)\;.
  \label{eqn:Iflux_correlated}
\end{equation}
Here, $D_{ij}$ refers to the combined covariance matrix defined in Eq.~\eqref{eqn:D_cov}.

It is instructive to consider two limiting cases.  If background uncertainties are negligible w.r.t.~Poisson noise, the first term in the right-hand side of Eq.~\eqref{eqn:D_cov} dominates, and we obtain the pure Poisson information flux
\begin{equation}
  \frac{d\mathcal{\widetilde F}_{11}}{dE} =
  \frac{I_1(E)^2}{I(E)}\;.
  \label{eqn:Iflux_pure}
\end{equation}
On the other hand, for a sufficiently large exposure, the second term in the right-hand side of Eq.~\eqref{eqn:D_cov} can dominate, and $D_{ij}$ becomes constant in time.  In that case, the effective information flux scales like $\propto \mathcal{E}^{-2}$, leading to a finite total measured information even for very large integration times.\footnote{This argument does not hold if the vector $x_i \equiv I_1/I_2(E_i)$ has components with zero eigenvalues w.r.t.~the matrix $M_{ij} \equiv \Sigma_\delta(E_i, E_j)$.  Components with zero eigenvalues correspond to characteristics of the signal that are completely uncorrelated with the modeled background uncertainties, and give rise to non-zero contributions to the effective information flux even after very large integration times. This often undesired behaviour can be removed by adding a small diagonal contribution to the background uncertainty $M$, which also improves the numerical stability of the matrix inversions.}

\medskip

\begin{figure}[h]
  \centering
  \includegraphics[width=0.6\linewidth]{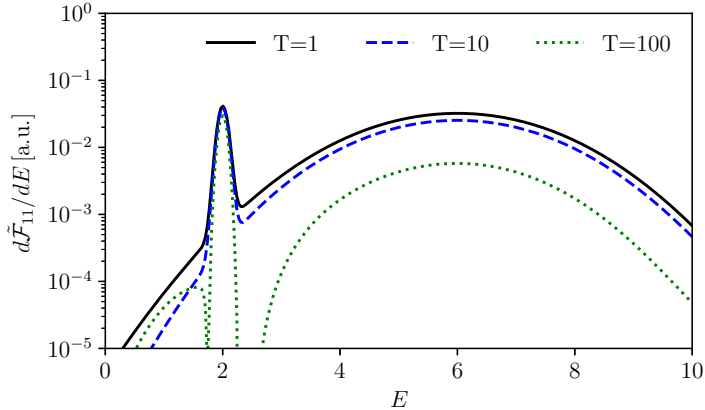}
  \caption{Example for the saturation effects of the effective information flux.  We show here the effective information flux for a signal with narrow and a broad spectral component, on top of a background with correlated uncertainties, after different observation times $\tobs$.  
  See Eq.~\eqref{eqn:Iflux_correlated} and text for further details.}
  \label{fig:dIdx}
\end{figure}

In Fig.~\ref{fig:dIdx} we show the effective information flux from Eq.~\eqref{eqn:Iflux_correlated}, for different values of the past observation time $\tobs$ (remember that $\mathcal{E}(E) = \aeff \tobs$, and we set $\aeff=1$).  For small observation times, the flux essentially corresponds to the pure Poisson contribution in Eq.~\eqref{eqn:Iflux_pure}.  However, for larger observation times, the information flux from the broad peak around $E=6$ becomes increasingly suppressed.  This is due to the fact that this broad feature is significantly degenerate with the modeled background uncertainties.  On the other hand, the flux from the narrow signal component around $E=2$, which has a width that is smaller than the correlation scale of the modeled background systematics, remains practically constant.  

\begin{figure}[h]
  \centering
  \includegraphics[width=0.6\linewidth]{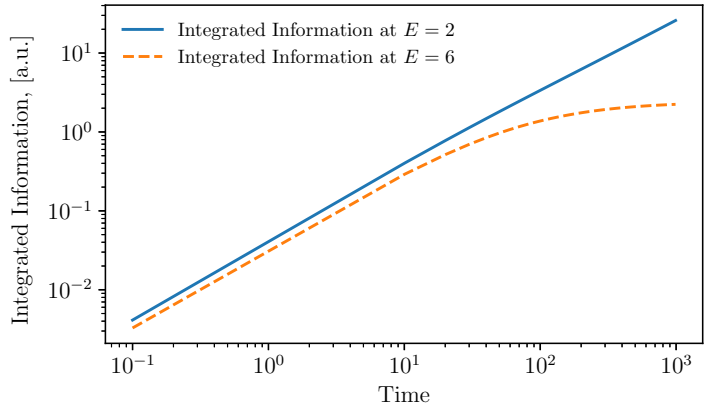}
  \caption{The cumulative information flux at two reference energies, corresponding to the scenario in Fig.~\ref{fig:dIdx}, but integrated over the observation time, as function of the total observation time $\tobs$.}
  \label{fig:dIdt}
\end{figure}

To further illustrate the saturation effects when measuring over a long time, we show in Fig.~\ref{fig:dIdt} the cumulative information obtained by integrating the effective information flux from Eq.~\eqref{eqn:Iflux_correlated} at the peaks of the two features, $E=2$ and $E=6$. It is clear that at $T=10^2$ the information obtained from observing the broad peak becomes saturated, while the sharper peak continues to provide information. 

\medskip

\paragraph*{Non-locality.} We demonstrate the non-locality of the effective information flux with a simple two-component example.  With non-locality, we mean that the information flux at $E$ depends in general on the past observation history of $E'\neq E$.  This makes sense, since usually a comparable exposure of different observational regions is required to break degeneracies between various model components.

\begin{figure}[h]
  \centering
  \includegraphics[width=0.6\linewidth]{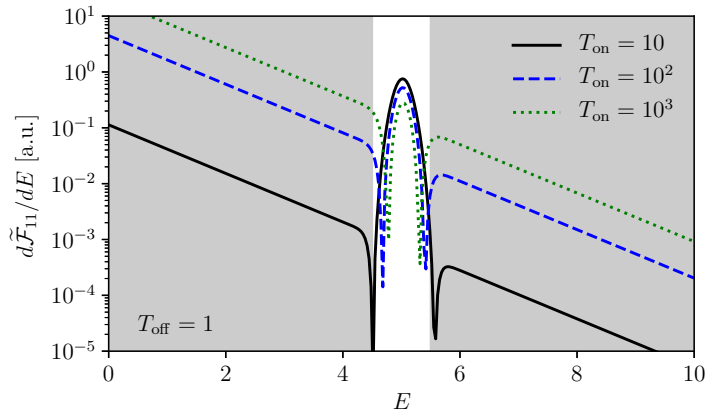}
  \caption{Example for the non-locality of the effective information flux.  We show the effective information flux for a narrow signal on top an exponential background with free normalization.  After an initial observation of the sidebands for $T_\text{off}=1$, observations are assumed to only take place in the narrow range $E=4.5$--$5.5$, and we show the effective information flux after different signal observation times $T_\text{on}$. See text for details.}
  \label{fig:dIdx2}
\end{figure}

In our simple two-component example, the signal flux, $I_1(E)$, is given by a normal distribution, centered at $E=5$, with a width of $\sigma=0.2$.  The background flux is given by $I_2(E) = 5\exp(-(E-5))$.  We assume that after an initial observation of the sidebands with $T_\text{off}=1$, observations only take place in the narrow range $E=4.5$--$5.5$.  We show the resulting effective information flux in Fig.~\ref{fig:dIdx2}.  At early observation times $T_\text{on}=10$, the information flux is completely dominated by observations of the signal region, since due to the initial sideband observations the background in the signal region is already reasonably constrained.  However, with growing observation time, the non-observations of the sidebands, becomes increasingly problematic.  Consequently, the information flux of the sidebands grows continuously.  

\section{Conclusions}
\label{sec:conclusions}

We introduced new methods and concepts for efficient and informative forecasting in astroparticle physics and DM searches, based on the powerful Fisher matrix formalism and unbinned Poisson likelihoods.  The Fisher information matrix, Eq.~\eqref{eqn:I}, is a way of quantifying the maximum information that an observation carries about a set of model parameters.  It is at the core of many more advanced statistical methods, and heavily used in other branches of science.

We introduced compact expressions for the approximate derivation of expected exclusion and discovery limits, Eqs.~\eqref{eqn:UL} and~\eqref{eqn:mlr} (\emph{equivalent counts method}).  The equivalent counts method is based on new definitions for the equivalent number of signal and background events, Eqs.~\eqref{eqn:si} and~\eqref{eqn:bi}.  These are solely based on the Fisher information matrix.  The equivalent counts method leads to surprisingly accurate results, even deeply in the Poisson regime, as we showed by comparison with the exact Neyman belt construction and maximum likelihood ratio techniques.  In this work, we focused on additive component models and Poisson processes, and neglected the effects of shape uncertainties.

We furthermore showed in two examples how systematic uncertainties can be efficiently accounted for within the Fisher formalism.  In the first example, we assumed that a component of the background is completely degenerate with the shape of the signal, Eq.~\eqref{eqn:sigma_stat_syst}.  In the second example, we modeled background uncertainties using the idea of Gaussian random fields with arbitrary correlation functions.  The resulting expressions, Eq.~\eqref{eqn:I_corr_syst}, can be efficiently solved numerically by matrix inversion, without profiling or marginalizing over the potentially thousands of nuisance parameters.

Finally, we introduced the new concept of \emph{information flux}, which we obtained from the Fisher information matrix by applying a functional derivative w.r.t.~the instrument exposure, Eq.~\eqref{eqn:Iflux}.  For unbinned Poisson likelihoods, it is equivalent to the commonly used signal-to-noise ratio.  However, when including the effect of nuisance parameters, the resulting \emph{effective information flux} accounts automatically for the non-local properties and saturation effects of background and instrumental uncertainties.  We illustrated these effects in two examples in Figs.~\ref{fig:dIdx} and~\ref{fig:dIdx2}.

Our motivation for this work was to provide a both solid and efficient statistical framework for the systematic study of optimal search strategies for a large range of dark matter models in indirect and other searches which will be the subject of future publications.  We furthermore plan to expand the discussion towards models with shape uncertainties, model discrimination and Fisher information geometry.

\medskip

In summary, we showed how to make the powerful Fisher matrix formalism useful for typical problems in astroparticle physics and DM searches.  The equivalent counts method for calculating expected exclusion and discovery limits is applicable in a large range of diverse situations, ranging from direct DM searches to the detectability of extended gamma-ray sources.  The effective Fisher information flux is a flexible tool for search strategy optimization, and we expect it to be particularly interesting when confronted with a large number of potential targets, like in indirect searches for DM or a large number of analysis channels in particle colliders.  In this work, we just scratched the surface of what can be done with the Fisher formalism, and expect fruitful further theoretical developments of the formalism in the future.

\acknowledgments

We thank Richard Bartels, Sebastian Liem and Christopher McCabe for useful discussions, and Eric Charles and Roberto Trotta for useful comments on the manuscript. This research is funded by NWO through an NWO VIDI research grant.

\bibliographystyle{JHEP}
\bibliography{stats.bib}

\appendix

\section{Poisson likelihood properties}
\label{apx:PL}

We discuss here the general Poisson likelihood function that we use in the main part of the paper, its higher-order derivatives, expectation values and skewness.

\subsection{Generalized Poisson likelihood}
\label{apx:GPL}

We concentrate in this section on univariate data (the generalization to the multivariate case is straightforward).  A typical example are photon energy spectra, measured over some fixed energy range, $E^- \dots E^+$. The data is then fully described by an unordered list of the energies of the $N$ measured photons,
\begin{equation}
  \mathcal{D} \equiv \{E_1, E_2, \dots, E_N\}\;.
\end{equation}
The corresponding PDF is

\begin{equation}
  P(\mathcal{D}|\vect\theta) =
  \frac{e^{-\mu(\vect \theta)}}{N!}\prod_{i=1}^N \Phi(E_i|\vect\theta)\;,
  \label{eqn:FullPoisson}
\end{equation}
where $\Phi(E|\vect\theta)$ is the model counts spectrum of the photons, and the expected total number of events can be calculated as
\begin{equation}
  \mu(\vect\theta) \equiv \int_{E^-}^{E^+} dE\; \Phi(E|\vect\theta)\;.
\end{equation}
The PDF in Eq.~\eqref{eqn:FullPoisson} is correctly normalized to one, which can be checked integrating over photon energies and summing over $N$.

The unbinned count map $\mathcal{C}(E)$, introduced in Eq.~\eqref{eqn:unbinned_countmap}, carries exactly the same information as the unordered photon list $\mathcal{D}$.  It turns out to be useful to rewrite the Poisson likelihood as function of the unbinned count map $\mathcal{C}(E)$.  A formal expression that is completely equivalent to Eq.~\eqref{eqn:FullPoisson} is
\begin{equation}
  P(\mathcal{C}|\vect\theta) = \exp\left( \int dE \left[\mathcal{C}\ln\Phi - \Phi\right] - \Gamma(N+1)
  \right)\;,
\end{equation}
where we used the gamma function instead of the factorial, $N!=\Gamma(N+1)$,  and the total number of measured events is given by
\begin{equation}
  N \equiv \int_{E^-}^{E^+} dE\,\mathcal{C}(E)\;.
\end{equation}
The key advantage is here that the domain on which $P(\mathcal{C}|\vect\theta)$ is defined can be immediately extended to arbitrary functions $\mathcal{C}(E)$, which can be continuous and/or feature non-integer total measured events numbers.  This will become useful below.

\medskip

The likelihood function corresponding to the Poisson process is given by
\begin{equation}
  \mathcal{L}(\mathcal{C}|\vect\theta) \equiv A(\mathcal{C})
  P(\mathcal{C}|\vect\theta)\;,
  \label{eqn:L}
\end{equation}
where $A(\mathcal{C})$ is an arbitrary positive function of the data that does not affect the discussion. The logarithm of Eq.~\eqref{eqn:L}, together with the choice $A(\mathcal{C}) = \exp(-\Gamma(N+1))$, leads Eq.~\eqref{eqn:lnLp} in the main text.

\subsection{Expectation values and `Asimov data'}
\label{apx:Asimov}

Writing the Poisson log-likelihood in the form of Eq.~\ref{eqn:lnLp} has the advantage that it is a linear function of the counts map $\mathcal{C}$.  This means that averages over projected data are trivial.  Remember that the unbinned count map averaged over many realizations of model $\vect\theta$ just equals the expected count map, $\langle \mathcal{C}\rangle_{\mathcal{D}(\vect\theta)} = \Phi(\vect\theta)$.  Hence, for linear functions of $\mathcal{C}$, averaging over data is equivalent to substituting the unbinned counts map by the `Asimov' data set~\cite{Cowan:2010js}, $\mathcal{C}\to \Phi(\vect\theta)$.  

Maximizing $\ln\mathcal{L}$ with respect to $\vect\theta$ requires that $\partial\ln\mathcal{L}/\partial\theta_k= 0$, which is equivalent to
\begin{equation}
  \int_{E^-}^{E^+} dE\, \left(
  \mathcal{C}(E)
  \frac{\Psi_k(E)}{\Phi(E|\vect{\theta})}
  -\Psi_k(E) \right)= 0\;,
\end{equation}
for all components $k$.  We adopted here the additive component model defined in
Eq.~\eqref{eqn:add_model}.  Replacing $\mathcal{C}$ by Asimov data for model $\vect\theta$ yields zero, as expected.

For the additive component model, averages over higher order derivatives of the log-likelihood take a simple form.  The $n$-th order derivative (for $n\geq2$) reads
\begin{equation}
  \left\langle\frac{\partial^n(-\ln\mathcal{L})}{\partial \theta_{k_1}\dots\partial\theta_{k_n}}\right\rangle_{\mathcal{D}(\vect\theta)}\\
  = (-1)^n(n-1)!\int_{E^-}^{E^+} dE\,
  \frac{\Psi_{k_1}(E)\dots\Psi_{k_n}(E)}{\Phi^{n-1}(E|\vect\theta)} \;.
\end{equation}
For $n=2$, we recover Eq.~\eqref{eqn:Ipsi} in the main text.  The expression for $n=3$ is important for the discussion of the non-Gaussianity effects in the next subsection.

\subsection{The Gaussian regime}
\label{apx:Gregime}

We start by Taylor expanding the log-likelihood around the model parameter $\vect\theta$,
\begin{equation}
  \ln\mathcal{L} = \text{const}+
  \sum_{i}\Delta\theta_i
  \frac{\partial\ln\mathcal{L}}{\partial \theta_i}
  +
  \frac{1}{2!}\sum_{ij}\Delta\theta_i\Delta\theta_j
  \frac{\partial^2\ln\mathcal{L}}{\partial \theta_i \partial \theta_j}
  +\\
  \frac{1}{3!}\sum_{ijk}\Delta\theta_i\Delta\theta_j\Delta\theta_k
  \frac{\partial^3\ln\mathcal{L}}{\partial\theta_i\partial\theta_j\partial\theta_k}
   +\dots\;,
\end{equation}
where $\Delta \vect\theta$ is the deviation from the expansion point.  If we average the $\ln\mathcal{L}$ now over model realizations $\mathcal{D}(\vect\theta)$, and assume Poisson likelihoods, this expansion can be written as

\begin{equation}
  \langle -\ln\mathcal{L}(\vect\theta+\Delta\vect\theta) \rangle_{\mathcal{D}(\vect\theta)} =
  \text{const} + \\
  \frac12\sum_{ij} \Delta\theta_i \Delta\theta_j
  \left( \mathcal{I}_{ij} + \frac{2}{3} \sum_k\Delta\theta_k
    \frac{\partial\mathcal{I}_{ij}}{\partial \theta_k}
  \right) + \dots\;.
  \label{eqn:lnL_expansion}
\end{equation}
Note that the additional factor two in front of
$\partial\mathcal{I}_{ij}/\partial\theta_k$ comes from the fact that the
derivative is here also affecting the parameters of the Asimov data.  The precise factor depends on the actual likelihood function and its dependence on the data.

\medskip

In order to understand the impact of higher-order terms in Eq.~\eqref{eqn:lnL_expansion}, it is convenient to think about how they affect the significance at the boundaries of confidence intervals.  Naively, the boundary of a confidence interval that extends to a threshold value $t$ in a certain direction would correspond to the ellipsoid constructed by values of $\Delta\theta_i$ that satisfy the equation
\begin{equation}
  \sum_{ij}\Delta\theta_i \Delta\theta_j \mathcal{I}_{ij} = t\;.
\end{equation}
When taking into account third-order terms from Eq.~\eqref{eqn:lnL_expansion},
the \emph{actually} realized threshold value at point $\Delta\theta_i$ on the ellipsoid changes by
\begin{equation}
  \Delta t = \frac23\sum_{ijk}\Delta\theta_i\Delta\theta_j\Delta\theta_k
  \frac{\partial\mathcal{I}_{ij}}{\partial\theta_k}\;.
\end{equation}
If $\Delta t \ll t$ at all points of the ellipsoid, higher-order (more
precisely third order) terms can be indeed ignored and do not affect the result.

In practice, it is usually sufficient to concentrate on the principal axes of the
ellipsoid, which correspond to the eigenvectors of $\mathcal{I}_{ij}$.  For simplicity, we will here further assume that there are no strong degeneracies between the components.  In that case, the principal axes approximately align with the individual parameters $\theta_i$.  We can then, for every direction $i$, require that
\begin{equation}
  \frac{\Delta t}{t} = \frac{2}{3} \frac{\Delta\theta_i}{\mathcal{I}_{ii}}
  \frac{\partial \mathcal{I}_{ii}}{\partial\theta_i} < f\;,
\end{equation}
which means that the fractional change in the threshold value $t$ should be
smaller than $f$. For the Poisson likelihood that we assumed already above in Eq.~\eqref{eqn:lnL_expansion}, we can use
$\Delta\theta_i(\mathcal{I}_{ii})^{-1} \partial
\mathcal{I}_{ii}/\partial\theta_i =
\Delta\theta_i\sqrt{\mathcal{I}_{ii}}/\sqrt{b_i} =  \sqrt{t/b_i}$, where we used the definition of the equivalent background in Eq.~\eqref{eqn:bi}. This implies the condition
\begin{equation}
  b_i > \frac{4t}{9f^2}\;,
\end{equation}
which is identical to Eq.~\eqref{eqn:b_cond} in the main text and further discussed there.

We emphasize that this is a quite naive estimate, and does not take into account the possible effects of parameter degeneracies, deviations of the log-likelihood ratio from a chi-square distribution, etc.  But it gives a useful heuristics for when the Fisher formalism is safe to use, and when it should be used with care.  As we have seen in Sec.~\ref{sec:estimates}, if the appropriate prescriptions are used, reasonable results for expected exclusion and discovery limits on \emph{one} signal component of interest can be obtained even for vanishing equivalent backgrounds, as long as the other components are well constrained and behaved.

\section{Expected exclusion and discovery limits}
\label{apx:UL}

We describe here very briefly some of the exact methods that are used to calculate expected exclusion and discovery limits.  These are in the main text compared against the results from our EC method.

\subsection{Neyman belt construction}

For details about the Neyman belt construction, we refer to Ref.~\cite{Feldman:1997qc}.  We repeat here just the technical result.  A conventional one-sided upper limit on the number of expected signal events, $s$, given $k$ observed events and $b$ expected background events, is given by the $s^U$ that satisfies the equation
\begin{equation}
  \sum_{k'\leq k} P(k'|s^U+b) = \alpha\;,
\end{equation}
where $\alpha$ is the significance level of the limit, and $P(c|\mu)$ the PDF of the Poisson distribution with $c$ observed and $\mu$ expected counts.  Since the connection between $k$ and $s^U$ is monotonic, the median expected exclusion limit can be obtained by considering, for a given expected background $b$, the median expected count number $k$.  Since $k$ can only acquire discrete values, this introduces jumps in the projected upper limits as function of $b$, which are clearly visible in Fig.~\ref{fig:UL1}.

\medskip

The minimum number of events $k$ that leads to a signal detection with a significance level of (at least) $\alpha$, above an expected background of $b$ events, is given by
the smallest $k$ that satisfies the inequality
\begin{equation}
  \sum_{k'\geq k} P(k'|b) \leq \alpha\;,
\end{equation}
which we call here $k_\text{th}$.  The expected discovery limit of the signal for the given sensitivity level is now given by the smallest $s$ that corresponds to a median count of $k_\text{th}$.

\subsection{Maximum likelihood ratio method}

For general likelihood functions with $n$ free parameters, we can define the MLR test statistic (see Ref.~\cite{Cowan:2010js} for details)
\begin{equation}
  TS_\mathcal{D}(\theta_1) = -2 \ln\frac
  {\max_{\theta'_2, \dots, \theta'_n\geq0} \mathcal{L}(\mathcal{D}|\theta_1, \theta'_2, \dots, \theta'_n)}
  {\max_{\theta'_1, \dots, \theta'_n\geq0} \mathcal{L}(\mathcal{D}|\theta'_1, \dots, \theta'_n)}\;.
\end{equation}
We fix here only one parameter, $\theta_1$, and maximize w.r.t.~the remaining ones.  We define the modified test statistic
\begin{equation}
  q_\mathcal{D}(\theta_1) = \left\{\begin{array}{lcl}
      TS_\mathcal{D}(\theta_1)  &&  \text{if} \quad \theta_1 > \hat\theta_1(\mathcal{D}) \\
      0  && \text{otherwise}
  \end{array}\right.\;,
\end{equation}
where $\hat\theta_1\geq0$ is the maximum likelihood estimator given data $\mathcal{D}$.  For a given data set, an one-sided confidence interval that corresponds to the desired upper limits is given by
\begin{equation}
  C = \{\theta_1 \geq 0\ |\ q_\mathcal{D}(\theta_1) \leq t(\theta_1) \}\;,
\end{equation}
where the threshold $t(\theta_1)$ is in general a function of $\theta_1$ and depends on the aspired significance level $\alpha$.  The threshold must be set such that $C$ has correct coverage properties.  This means that $C$ should cover the true value of $\theta_1$ in $1-\alpha$ of the cases.  In the large-sample limit, asymptotic formulae for the statistical distribution of $q_\mathcal{D}(\theta_1)$ are available~\cite{Cowan:2010js}.  However, in the small-sample regime, MC simulations are required to derive appropriate threshold values for $t(\theta_1)$.  Note that $C$ can be the empty set in some cases, which corresponds to downward fluctuations of the background.  Although there are numerous ways to deal with this situations~\cite{Rolke:2004mj, Feldman:1997qc}, this is not problematic for the purposes of the present work.  Note that, due to the discreteness of Poisson processes, $t(\theta_1)$ is in general not a continuous function of $\theta_1$ or the other background parameters.

In the main text, we usually show median limits obtained from a large set of data realizations with $\theta_1=0$, using a $t(\theta_1)$ that is derived from MC simulations.

\medskip

Expected discovery limits are derived in a similar way.  We first find the threshold value $t_\text{th}$ that corresponds to a test of the hypothesis $\theta_1=0$ with the significance level of $\alpha$,
\begin{equation}
  P(TS_\mathcal{D}(\theta_1=0) \geq t_\text{th}|\theta_1 = 0) \leq \alpha\;.
\end{equation}
The discovery limit is then given by the smallest value of $\theta_1$ that leads to a detection in at least $50\%$ of the cases, namely we search for the $\theta_1^D$ that satisfies
\begin{equation}
  P(TS_\mathcal{D}(\theta_1=0) \geq t_\text{th}|\theta_1 = \theta_1^D) \geq 0.5\;.
\end{equation}
Note that $t_\text{th}$ and hence $\theta_1^D$ are not necessarily smooth functions of the background parameters $\theta_2, \dots, \theta_n$, if the discreteness of the Poisson likelihood plays a role.

\section{Technical calculations}
\label{apx:technical}

We present here some more details about derivations of equations related to
discovery limits as well as the treatment of background systematics, used
in the main part of the paper.

\subsection{Expected discovery limits}
\label{apx:gammaD}

Given $b\ll1$ expected background events, one can derive an approximate discovery limit for the number of required signal events $s$ and statistical significance $\alpha$, by solving
\begin{equation}
  P(s+b|b) = \frac{e^{-b} b^{s+b}}{\Gamma(s+b+1)} =
  \alpha\;,
\end{equation}
for $s$.  Here, $P(s+b|b)$ is the continuum version of the Poisson probability mass function, with $s+b$ observed events while $b$ are expected.

We compare this expression with Eq.~\eqref{eqn:mlr} (we use the symbols $s$ and $b$ for simplicity).  In the limit $b\ll s$ it can be written as
\begin{equation}
  \label{eqn:mlr2}
  s\ln \left(\frac{s}{b}\right)
  - s = \frac{Z^2}{2}\;.
\end{equation}
Now, one can consider the first two terms of the expansion of $\alpha$ in $1/Z$,
\begin{equation}
  \ln\frac1\alpha \simeq \frac{Z^2}{2} + \frac12\ln2\pi Z^2\;,
\end{equation}
which can be substituted into the right-hand side of Eq.~\eqref{eqn:mlr2}.  The large-$s$ approximation to the log of the gamma function, Stirling's formula, reads (we use $b\ll s$)
\begin{equation}
  \ln \Gamma(s+1) \approx s\ln s - s + \frac12 \ln 2\pi s\;,
\end{equation}
which can be substituted in the left-hand side of Eq.~\eqref{eqn:mlr2}. One can then rearrange the terms such that they read
\begin{equation}
  \frac{b^s}
  {\Gamma(s+1)}
  = \alpha \cdot \sqrt{\frac{Z^2}{s}}\;.
\end{equation}
This has exactly the form shown in Eq.~\eqref{eqn:poisAprox} (remember that $\lambda_i \simeq s_i$ in the low-background limit).

\subsection{Examples with systematic errors}
\label{apx:syst_errors}

We will show here in some detail how to arrive at the results in
Eqs.~\eqref{eqn:sigma_stat_syst} and~\eqref{eqn:I_corr_syst}.

\medskip

In the example leading to Eq.~\eqref{eqn:sigma_stat_syst}, we have a three
component system with a signal, a background, and some third component which
parametrizes variations in the background that are completely degenerate with
the signal.  The full Fisher matrix of the system is given by
\begin{equation}
  \mathcal{I} =
  \left(
    \begin{array}{ccc}
      \mathcal{I}_{11}^\text{pois} & \mathcal{I}_{12}^\text{pois} & \mathcal{I}_{11}^\text{pois} \\
      \mathcal{I}_{12}^\text{pois} & \mathcal{I}_{22}^\text{pois} & \mathcal{I}_{12}^\text{pois} \\
      \mathcal{I}_{11}^\text{pois} & \mathcal{I}_{12}^\text{pois} & \mathcal{I}_{11}^\text{pois}+\frac{1}{\xi_3^2} \\
    \end{array}
  \right)\;,
\end{equation}
where we already used the various symmetry properties of the Fisher information matrix elements as well as the fact that $\Psi_1 = \Psi_3$.  We are interested in the profiled $1\times1$ Fisher matrix where the parameter of interest is the signal parameter $\theta_1$, and we have removed $\theta_2$ and $\theta_3$.  This profiled Fisher matrix is given by
\begin{equation}
  \mathcal{\widetilde I} =
  \mathcal{I}_{11}^\text{pois}
  \\-
  \left(
    \begin{array}{ccc}
      \mathcal{I}_{12}^\text{pois} \\
      \mathcal{I}_{11}^\text{pois}
    \end{array}
  \right)^T
  \left(
    \begin{array}{ccc}
      \mathcal{I}_{22}^\text{pois} & \mathcal{I}_{12}^\text{pois} \\
      \mathcal{I}_{12}^\text{pois} & \mathcal{I}_{11}^\text{pois}+\frac{1}{\xi_3^2} \\
    \end{array}
  \right)^{-1}
  \left(
    \begin{array}{ccc}
      \mathcal{I}_{12}^\text{pois} \\
      \mathcal{I}_{11}^\text{pois}
    \end{array}
  \right)\;.
\end{equation}
It is straightforward to invert the $2\times2$ matrix analytically, and one can show that the inverse of the profiled Fisher matrix can be written in the simple form
\begin{equation}
  \frac{1}{\mathcal{\widetilde I}}=
  \xi_3^2 +
  \frac1{
    \mathcal{I}_{11}^\text{pois}
    -
    (\mathcal{I}_{22}^\text{pois})^{-1}
    (\mathcal{I}_{12}^\text{pois})^2
  }\;.
\end{equation}
If we now identify $\sigma_1^2(\vect\theta) = \mathcal{\widetilde I}^{-1}$, and $(\sigma_1^\text{pois})^2(\vect\theta) = (\mathcal{I}_{11}^\text{pois} - (\mathcal{I}_{22}^\text{pois})^{-1} (\mathcal{I}_{12}^\text{pois})^2)^{-1}$, we arrive at Eq.~\eqref{eqn:sigma_stat_syst}.  Note that the latter is just the profiled Fisher information that we would have obtained in absence of the third component, or equivalently in the limit $\xi_3\to\infty$.

\medskip

The derivation of Eq.~\eqref{eqn:I_corr_syst} follows a similar pattern, but is technically slightly more involved.  Again, we are interested in the profiled $1\times1$ Fisher information for the signal component.   It is here useful to associate the index $i=0$ with the signal component, and the indices $i=1, \dots, N$ with the discrete energies from Eq.~\eqref{eqn:Delta}.  If we think about the underlying full Fisher matrix of the system in the block form shown in Eq.~\eqref{eqn:block}, then the components $A$ (associated with $i=0$), $B$ and $C$ are given by
\begin{equation}
  A = \sum_{i=1}^N \Delta E_i \frac{\Psi_1(E_i)^2}{\Phi(E_i)}\;,
\end{equation}
where we discretized the integral,
\begin{equation}
  C_i = \Delta E_i \frac{\Psi_1(E_i)\Psi_2(E_i)}{\Phi(E_i)} =
  \Delta E_i \Psi_1(E_i)\;,
\end{equation}
where $i = 1, \dots, N$ and we used in the second step that $\Phi(E) = \Psi_2(E)$, and
\begin{equation}
  B = \delta_{ij} \Delta E_i \Phi(E_i) + \Sigma_\delta^{-1}\;,
\end{equation}
where we included the inverse of the covariance matrix for $\xi_i$.  Then, the profiled Fisher information can be written as
\begin{multline}
\begin{split}
  \mathcal{\widetilde I}_{11} &= \sum_{i=1}^N \Delta E_i \frac{\Psi_1(E_i)^2}{\Phi(E_i)}
  -\sum_{i,j=1}^N
  \Delta E_i \Delta E_j \Psi_1(E_i) \Psi_1(E_j)
  \left[
    \text{diag}(\Delta E_i \Psi_2(E_i)) + \vect\Sigma_\delta^{-1}
  \right]^{-1}_{ij}\\
  &= \sum_{i=1}^N \Delta E_i \frac{\Psi_1(E_i)^2}{\Phi(E_i)}
  -\sum_{i,j=1}^N
  \sqrt\frac{\Delta E_i \Delta E_j}{\Phi(E_i)\Phi(E_j)} \Psi_1(E_i) \Psi_1(E_j)
  \left[\frac{\vect\Sigma'_\delta}{1+\vect\Sigma'_\delta}\right]_{ij}\\
  &= \sum_{i,j=1}^N
  \sqrt\frac{\Delta E_i \Delta E_j}{\Phi(E_i)\Phi(E_j)} \Psi_1(E_i) \Psi_1(E_j)
  \left[\frac{1}{1+\vect\Sigma'_\delta}\right]_{ij}\;.
\end{split}
\end{multline}
Here, we used the definitions $(\vect\Sigma_\delta)_{ij} = \Sigma_\delta(E_i,
E_j)$ and $(\vect\Sigma'_\delta)_{ij} = \sqrt{\Delta E_i \Delta E_j \Phi(E_i) \Phi(E_j)}\,\Sigma_\delta(E_i, E_j)$.  In the first step, we rearrange some factors of $\Delta E_i$ and $\Phi_1(E_i)$, and use the general matrix relation $(1+M^{-1})^{-1} = M/(1+M)$.  In the second step, we include the matrix identity in the form $\delta_{ij} = [(1+\vect\Sigma'_\delta)/(1+\vect\Sigma'_\delta)]_{ij}$ in the first summation which helps to further collapse the whole expression.  The last line is after some more rewriting equivalent to Eq.~\eqref{eqn:I_corr_syst}.

\end{document}